\documentclass[11pt]{article}
\usepackage[sc]{mathpazo} 
\usepackage{fullpage}
\usepackage[authoryear,sectionbib,sort]{natbib}
\usepackage[utf8]{inputenc}
\usepackage{lineno}
\usepackage{titlesec}
\usepackage{graphicx,amsmath}
\usepackage{textcomp}

\titleformat{\section}[block]{\Large\bfseries\filcenter}{\thesection}{1em}{}
\titleformat{\subsection}[block]{\Large\itshape\filcenter}{\thesubsection}{1em}{}
\titleformat{\subsubsection}[block]{\large\itshape}{\thesubsubsection}{1em}{}
\titleformat{\paragraph}[runin]{\itshape}{\theparagraph}{1em}{}[. ]

\title{Evolution at the edge of expanding populations}

\author{Maxime Deforet$^{1,\ast}$ \\ 
Carlos Carmona-Fontaine$^{2}$ \\ 
Kirill S. Korolev$^{3}$ \\
Joao B. Xavier$^{4}$}

\date{}

\begin{document}

\maketitle

\noindent{} 1. Sorbonne Université, Centre National de la Recherche Rcientifique, Laboratoire Jean Perrin, LJP, Paris 75005, France;

\noindent{} 2. Center for Genomics and Systems Biology, Department of Biology, New York University, New York City, New York 10003;

\noindent{} 3. Department of Physics and Graduate Program in Bioinformatics, Boston University, Boston, Massachusetts 02215;

\noindent{} 4. Program in Computational Biology, Memorial Sloan-Kettering Cancer Center, New York City, New York 10065.

\noindent{} $\ast$ Corresponding author; e-mail: maxime.deforet@sorbonne-universite.fr.

\bigskip

\textit{Keywords}: Range expansion, trade-off, dispersal, evolution, reaction-diffusion, \textit{Pseudomonas aeruginosa}.

\bigskip

\bigskip

\newpage{}

\section*{Abstract}

Predicting evolution of expanding populations is critical to control biological threats such as invasive species and cancer metastasis. Expansion is primarily driven by reproduction and dispersal, but nature abounds with examples of evolution where organisms pay a reproductive cost to disperse faster. When does selection favor this ‘survival of the fastest?’ We searched for a simple rule, motivated by evolution experiments where swarming bacteria evolved into an hyperswarmer mutant which disperses $ \sim 100\%$ faster but pays a growth cost of $\sim 10 \%$ to make many copies of its flagellum. We analyzed a two-species model based on the Fisher equation to explain this observation: the population expansion rate ($v$) results from an interplay of growth ($r$) and dispersal ($D$) and is independent of the carrying capacity: $v=2\sqrt{rD}$. A mutant can take over the edge only if its expansion rate ($v_2$) exceeds the expansion rate of the established species’ ($v_1$); this simple condition ($v_2 > v_1$) determines the maximum cost in slower growth that a faster mutant can pay and still be able to take over. Numerical simulations and time-course experiments where we tracked evolution by imaging bacteria suggest that our findings are general: less favorable conditions delay but do not entirely prevent the success of the fastest. Thus, the expansion rate defines a traveling wave fitness, which could be combined with trade-offs to predict evolution of expanding populations.

\newpage{}

\section*{Introduction}

Biological threats often come in the form of expanding populations: A cancerous tumor spreads into a healthy tissue; bacteria colonize a clean surface and form a biofilm; exotic species occupy a new territory. Predicting the evolution of expanding populations, however, is a complex problem. Expansion can be a combination of many organismal traits, so evolutionary trajectories can occur in a multi-dimensional phenotypic space.

For the sake of simplicity, we can reduce phenotype into two traits: dispersal and growth. Individuals move and they consume local resources; resource availability is highest outside the population range, which creates an advantage to being at the population margin (\citealt{Murray2007}). Therefore, there are two possible favorable evolutionary strategies: dispersing faster or growing faster. Fast-dispersing individuals take advantage of this spatial heterogeneity: they take over the edge, cutting-off competitors’ access to growth-limiting resources (\citealt{Nadell2010,Phillips2010}). In contrast, faster-growth individuals outcompete the rest of the population regardless of their location. Of course, simultaneously improving both traits—dispersal \textit{and} growth—is even better. It is more delicate, and perhaps more interesting, to predict what could happen when one trait is improved at the expense of the other, which is often the case if organisms live with limited resources.  For instance, if a mutant appears with better dispersal but has a lower growth rate because it spends too much energy on moving, will this mutant take over the population by reaching the edge, or will it be out-competed by the faster growing but slower dispersing wild-type? 

There are many examples suggesting that population expansion selects for better dispersal, even at the cost of slower growth (\citealt{Chuang2016}). The invasion of the cane toads in Australia, a human-introduced species, is led by faster long-legged individuals with lower birth rates (\citealt{Hudson2015}); the South African mountain fynbos is threatened by invasive pine trees with lighter pine seeds that disperse better (\citealt{Richardson1990}) but produce weaker seedlings (\citealt{Reich1994}); metastatic cancer cells are more invasive due to a loss of contact inhibition of locomotion (\citealt{Carmona-Fontaine2008}) that also lowers their cellular proliferation rates (\citealt{Biddle2011, Gerlee2012, Kim2017, Widmer2012}). Additional field examples of invasive populations, where margin individuals acquired greater dispersal and slower growth, include other plants (\citealt{Ganeshaiah1991, Huang2015, Williams2016}), fish (\citealt{Agostinho2015}), crickets (\citealt{Simmons2004}), butterflies (\citealt{Hughes2003}), and fungi (\citealt{Garbelotto2015}). Laboratory experiments with populations expanding towards a virgin territory with freshwater ciliates (\citealt{Fronhofer2015}), beetles (\citealt{Ochocki2017, Weiss-Lehman2017}), plants (\citealt{Williams2016}), and bacteria (\citealt{Fraebel2017, Ni2017}) led to similar results: population expansion can favor faster dispersal at the expense of slower growth.

Yet, previously proposed models suggest that faster growth is not always selected for. Growth can be traded off with competitive ability as in the r-K selection theory (\citealt{Pianka1970}) and, in a spatially structured environment, the competition-colonization trade-off theory aims to explain the coexistence of interacting species (\citealt{Tilman1994}). Nonetheless, these findings suggest that a better definition of fitness is required to understand evolution in expanding populations. Other questions ensue: Are there general conditions for favoring dispersal over growth? And how much cost can a fast-dispersing individual pay in terms of slower growth and still be favored by natural selection?

Here, we based our analysis on a well-established framework of spatial expansion in growing populations: the traveling wave derived from the Fisher-Kolmogorov-Petrovsky-Piscunov (F-KPP) equation. The F-KPP equation, in its original form, describes a 1-D monospecies population (\citealt{Fisher1937, Kolmogorov1937, Giometto2014}). We expanded the F-KPP equation to investigate the conditions favoring faster dispersal or faster growth rate, and we solved the resulting two-species system to produce a simple rule governing the evolutionary outcome. Somewhat surprisingly, this rule had not been proposed before to the best of our knowledge, despite much theoretical and experimental work in this field. We then conducted simulations to delineate the conditions at which the rule is applicable, and the time-scales necessary for a full sweep of the population in biologically relevant situations. This rule allowed us to calculate the maximum cost in term of growth rate that a faster-dispersal mutant can pay and still win the competition. If the loss of growth rate is greater than this maximum cost, then better dispersal should no longer be favored. When a physiological trade-off between growth and dispersal is considered as well, then it is possible to predict the phenotype favored by natural selection.

It is often challenging to test the predictions of theoretical models with field studies, and experimental manipulation of natural ecosystems is often impractical. But we can use laboratory experiments with microbes to rigorously test our mathematical models (\citealt{Dai2013, Gandhi2016, Hallatschek2007, Jessup2004, Mitri2016}). Microbial model have the advantages of large populations sizes, short generation times, affordable DNA sequencing and—in many cases—tools for genetic engineering. We recently discovered that experimental evolution in swarming colonies of the bacterium \textit{Pseudomonas aeruginosa} leads to the spontaneous evolution of hyperswarmers (\citealt{Ditmarsch2013}). We used DNA sequencing and genetic engineering to show that hyperswarmer mutants have a single point mutation in a gene called \textit{fleN}, which gives them multiple flagella and makes them more dispersive, and we confirmed that this evolution is reproducible in dozens of replicate experiments. Importantly, the many flagella always came at the cost of a slower growth (Table 1). \textit{P. aeruginosa} wild-type individuals outcompete hyperswarmers in well-mixed liquid media where faster dispersal is useless; hyperswarmers, on the other hand, swarm faster on agar gel (\citealt{Deforet2014}) and outcompete the wild-type in this spatially structured environment where dispersal is key (\citealt{Ditmarsch2013}). Thus, the hyperswarmer-wild-type dynamics can be used as a laboratory model to study the evolution in expanding populations where faster better dispersal comes with a growth cost. 

Here we exploited the differences in growth rate and dispersal between the wild-type \textit{P. aeruginosa} and its hyperswarmer mutant to experimentally test our model using time-course experiments with bacteria engineered to express fluorescent labels. The quantitative experiments supported our model, suggesting that the theory—despite its simplicity—provides a general way to predict the evolution of expanding populations in a range of biological species and systems.

\section*{Methods}

\subsection*{Theoretical model}

\subsubsection*{One dimensional one species F-KPP}
We modeled \textit{P. aeruginosa} swarming population as a clonally reproducing population, expanding along a one-dimension axis towards an open habitat, according to the F-KPP equation:
\begin{equation}
\frac{\partial u}{\partial t}=ru(1-\frac{u}{K})+ D \frac{\partial^2u}{\partial x^2} 
\label{eq:fkpp}
\end{equation}
where $x$ is space, $t$ is time, $u$ is the local population density, $K$ is the carrying capacity, $r$ is the maximum per-capita growth rate and $D$ quantifies dispersal. Growth and dispersal can obey different laws in nature; for generality, the F-KPP equation assumes logistic growth, where the per-capita growth rate decreases linearly as the population density increases, and assumes Fickian diffusion for dispersal. The F-KPP equation suits the common scenario where regions with excess of nutrients lie outside the population and determine the direction of expansion. Resource availability, proxied by $1-u/K$, is highest outside the population range; per capita growth, represented by $r(1-u/k)$, is maximal at the edge of the population. Eq. 1 has a traveling wave solution, $u(x,t)=u_0(x-vt)$, where the population front travels at a constant expansion rate $v=2\sqrt{rD}$, independent of the carrying capacity, and its density increases from the edge with a length-scale $\lambda=\sqrt{D/r}$ (Video 1 and Fig. \ref{fig:figS1}) (\citealt{Hallatschek2008, Murray2007}).

\subsubsection*{Edge of the population}
The population density decays exponentially at the front. The range of the traveling wave is theoretically infinite. Therefore, in order to locate the front position, we arbitrarily defined the “edge” as the location where the density reaches $5\%$ of the carrying capacity.

\subsubsection*{Two species F-KPP}
The F-KPP equation is extended to a two-species system with coupled equations:
\begin{equation}
\begin{aligned}
\frac{\partial u_1}{\partial t}&=r_1 u_1(1-u_1 - u_2)+ D_1 \frac{\partial^2u_1}{\partial x^2} \\
\frac{\partial u_2}{\partial t}&=r_2 u_2(1-u_1 - u_2)+ D_2 \frac{\partial^2u_2}{\partial x^2}
\end{aligned}
\end{equation}
Species 1 has density function $u_1(x,t)$, disperses with coefficient $D_1$ and grows with a rate $r_1$; species 2 has $u_2(x,t)$, $D_2$, and $r_2$. Species 1 and 2 interact only by competing for the same resources, a feature implemented by the factor $1-u_1-u_2$.

\subsubsection*{Competition}
In competition situations, we define the winning species as the resident species at the edge, namely the species whose frequency exceeds $50\%$ at the edge of the population (defined at the location where the total population becomes lower than $5\%$ of the carrying capacity). 

\subsubsection*{Numerical simulations}
The deterministic numerical simulations (used for Fig. \ref{fig:fig1}B, Fig. \ref{fig:fig2}C, and Fig. \ref{fig:fig3}A) were performed in MATLAB (The MathWorks) following Euler’s method, with $dx=0.1$, $dt=0.001$, $D_1 = 1$, $r_1 = 1$, and total spatial range of 400. For stochastic simulations (used in Fig. \ref{fig:fig2}B), the model was expanded as explained in Appendix C: Stochastic Modeling. 

\subsection*{Experimental Methods}
We used swarming motility in \textit{P. aeruginosa} as a laboratory model to study dynamics of expanding populations. Swarming plates (such as the one used for Fig. \ref{fig:fig1}A) were made as previously described (\citealt{Xavier2011}). They consist of soft agar gel supplemented with casamino acids and salts.

\subsubsection*{Transplantation experiments}
\textit{P. aeruginosa} strain PA14 genetically modified to constitutively express DsRed proteins were grown in LB overnight, washed twice in Phosphate Buffered Saline (PBS), then diluted in PBS to OD600=0.01. Each plate is seeded with 2 $\mu$L of bacterial solution and kept at 37\textdegree C for 20h. An overnight culture of hyperswarmers (clone 4) (\citealt{Ditmarsch2013}) genetically modified to constitutively express GFP proteins was washed twice in PBS and concentrated 100-fold by centrifugation. For each plate, the location of the tip of every branch of the colony was marked on the bottom side of the Petri dish. The implant sites were marked as well. A small volume of hyperswarmers culture (0.1-0.8 $\mu$L) was implanted at each implant site. From 6 to 11 branches were implanted per swarming colony. The entire procedure took less than 5 minutes per plate, which means the colony did not move significantly during the process. Immediately after implantation, each plate was placed inside a 37\textdegree C incubator containing a custom-made fluorescence imaging device. Two images were taken with the same light source (Blue LED equipped with a 500nm excitation filter): one with a 510nm emission filter (GFP channel), one without emission filter (brightfield channel). The camera dark noise and illumination unevenness were canceled out using this formula: 
\[
\text{Final image} = \frac{\text{GFP} – \text{dark noise}}{\text{brightfield} – \text{dark noise}}
\]
The size of each implant was manually evaluated from the total GFP signal within a region defined by thresholding. In order to make the experimental results comparable with simulations, this size was divided by the area of a circle of diameter of $\lambda_{\text{WT}}$. This gives the density of GFP as if the implant sites were $\lambda_{\text{WT}}$ in diameter. Then we divided this density by the wild-type carrying capacity. To evaluate the local carrying capacity of the wild-type colony ($K_{\text{exp}}$), we grew a swarming colony with a wild-type mutant constitutively expressing GFP proteins, took an image using the same imaging device and performing the same post-acquisition treatment, and measured the average intensity of the branches. Six hours after implantation, plates were imaged with a plate scanner (GE Healthcare Typhoon) in DsRed and GFP channels. The distance between the implant site and the location of the tip of the branch at the time of implantation was measured with ImageJ. The outcome was estimated visually. The amount of hyperswarmers at the front of the branch was visually ranked in four levels: (i) "No trace": no visible trace of hyperswarmers at the edge of the colony (red dots in Fig. \ref{fig:fig3}); (ii) "A few traces": a streak of hyperswarmers reached the edge (orange dots in Fig. \ref{fig:fig3}); (iii) "Partial sweep": hyperswarmers settled at the tip and along the edge of the branch (yellow dots in Fig. \ref{fig:fig3}); (iv) "Full sweep": hyperswarmers took over the ancestral population and disrupted the branch pattern (green dots in Fig. \ref{fig:fig3}). 

\subsubsection*{Growth curves}
Overnight cultures of wild-type and hyperswarmer cells were washed in PBS and diluted in minimum media with casamino acids (it is the same recipe as the one used for swarming plates except agar is removed). Cells were grown in a 96-well plate in a plate scanner (Tecan) with 37\textdegree C incubation and agitation. 

\subsubsection*{Competition experiments} 
Overnight cultures of wild-type DsRed and hyperswarmer GFP cells were washed in PBS and mixed to an approximate 1:1 ratio. To evaluate the pre-competition ratio, a sample of this mix solution was serially diluted in PBS and inoculated on a minimum media hard agar plate for CFU counting. 1 mL of the mix solution was poured on a fresh swarming plate. Once the plate was dry, it was incubated at 37\textdegree C for 4 hours. Finally, to evaluate the post-competition ratio, a small sample of the gel was scooped out using the wide end of a 1 mL sterile pipette tip to punch through the gel. The sample was placed in an Eppendorf tube with 0.5 mL of PBS, pipetted up and down 10 times to break the agar gel apart, vortexed for 10 seconds, then serially diluted in PBS and inoculated on a minimum media hard gar plate for CFU counting. CFU plates were scanned 24 hours later on a flatbed fluorescence scanner (Typhoon, GE Healthcare). Three competition plates per color combination were made per day (technical replicates). This experiment was performed three times (biological replicates). 

\subsubsection*{Data Availability} 
All experimental data (shown in Fig. \ref{fig:fig1} and Fig. \ref{fig:fig3}), and simulation results for Fig. \ref{fig:fig1} and Fig. \ref{fig:fig2}  (as well as MATLAB scripts to generate them), are available in the Dryad Digital Repository (\citealt{Dryad_Deforet}).

\section*{Results}

\subsection*{Modeling swarming in \textit{P. aeruginosa} with the F-KPP equation}

\textit{P. aeruginosa} populations swarm across agar gels containing nutrients and form branched colonies. Bacterial populations at the branch tips spread at a nearly constant rate (Table 1) by dividing and dispersing (\citealt{Deforet2014}). Knowing that cell sizes have a positive correlation with growth rates (\citealt{Deforet2015}) we compared the sizes of cells collected from the tip of a branch with the sizes of cells collected behind the tip; cells at the tip were longer, indicating faster growth at the edge of the population (Fig. \ref{fig:figS1}). Each growing tip consumes resources in its vicinity and thus forms a nutrient gradient (\citealt{Mitri2016}) that drives a resource-limited growth similar to the F-KPP model.

\subsection*{A simple rule for the evolution of faster dispersal}
Hyperswarmers grow $\sim 10\%$ slower in well-mixed liquid media due the cost of synthesizing and operating multiple flagella (Table 1), but, thanks to their $\sim 100\%$ faster dispersal on agar gel, they can outcompete the wild-type in spatially structured environments (\citealt{Deforet2014, Ditmarsch2013}). On agar gel lacking spatial structure, hyperswarmers are outcompeted, as expected (Fig. \ref{fig:figS2}). At the micrometer scale, an expanding population of hyperswarmers displays patterns of active turbulence typical of dense bacterial suspensions, which is different from the wild-type where cells remain nearly static even at the tips of swarming tendrils (Video 2).

\begin{figure}[h!]
	\includegraphics[width=14cm]{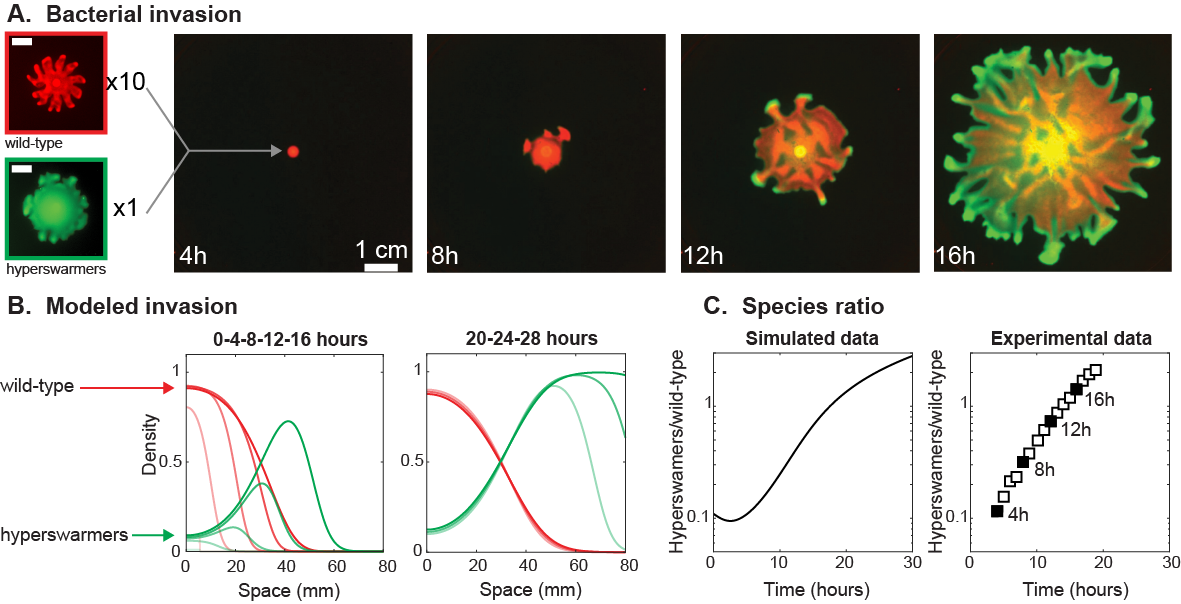}
	\caption{Evolutionary dynamics in an expanding population of swarming \textit{P. aeruginosa} bacteria. A: Fluorescence time-lapse imaging shows a swarming competition in a mixed population of wild-type and hyperswarmers at 10:1 ratio and on a soft-agar gel. Leftmost panels: Monoclonal swarming colonies of wild-type (top) and hyperswarmer (bottom) imaged at $T=12\text{h}$ (scale bars 1 cm). B: Snapshots of numerical simulations of Eq. 2 modeling competition of \textit{P. aeruginosa} (red lines) and the hyperswarmer mutant (green lines) with initial ratio 10:1, using parameters extracted from experiments (Table 1). Left panel shows time points represented in panel A. Right panel shows later time points, where the wild-type population stalls while the hyperswarmer population keeps expanding. C: Ratio of hyperswarmer biomass over wild-type biomass. The experimental data is extracted from fluorescence signals in panel A (black squares represent time points shown in A).}
	\label{fig:fig1}
\end{figure}

To gain a better understanding of the competition dynamics in expanding swarming colonies we mixed wild-type bacteria (labeled with the red fluorescent protein DsRed Express) with hyperswarmers (labeled with the green fluorescent protein GFP) at 10:1 ratio. We then used time-lapsed florescence imaging to film the swarming competition (Fig. \ref{fig:fig1}A). The time-lapse showed that hyperswarmers quickly reached the population edge, increasing their dominance as the colony expanded to win the competition (Video 3).

To determine the conditions favoring evolution of faster dispersal, we used an extension of the F-KPP equation for a two-species system, where $u_1$ represents the wild type and $u_2$ represents the hyperswarmer (See methods and models section). For simplicity, and according to data for the hyperswarmer system (\citealt{Ditmarsch2013}) (Fig. \ref{fig:figS3}), we assumed that both species have the same carrying capacity, which we normalized to 1. We assumed that their dispersal rates, determined by $D_1$ and $D_2$, are independent. This framework is well established and it has been used before to investigate competition in various contexts of range expansion (\citealt{Lewis2002, Okubo1989, Pigolotti2013}), including in a scenario with a linear trade-off between dispersal and growth (\citealt{Reiter2014}). It has been used also as a basis to elaborate more complex models (\citealt{Benichou2012, Bouin2012, Gandhi2016, Holzer2014, Guo2014, Lehe2012, Perkins2016, Ramanantoanina2014, King2003}). However, previous studies did not continue to derive a general rule for the evolutionary outcome of all possible values of dispersal and growth. 

To derive a general rule, we first investigated the conditions that allow an introduced population to thrive and replace the resident population at the expansion front. We could determine analytically that, in the moving reference frame traveling at the speed $v_1$, the frequency of species 2 at the edge grows at the rate 
\begin{equation}
\tilde{r} = \frac{v_2^2-v_1^2}{4D_2}
\label{eq:relativerate}
\end{equation}
which defines the relative fitness of species 2 within a population of species 1 (see Appendix A: Analytical solution for the condition of success). Species 2 outcompetes species 1 at the edge only if the relative fitness of species 2 is positive, which corresponds to:
\begin{equation}
v_2 > v_1
\label{eq:rule}
\end{equation}
where $v_1=2\sqrt{r_1 D_1}$ and $v_2=2\sqrt{r_2 D_2}$ are the expansion rates of each species when grown alone. 

Eq. 4 sets the conditions for success at the expansion edge. The intuition behind these evolutionary dynamics is well illustrated in a simulation of the competition between an established species (species 1) and a species with faster dispersal but slower growth (species 2), which we simulated (Fig. \ref{fig:fig1}B) by numerically solving the system in Eq. 2 with parameters corresponding to the hyperswarmer system ($r_2/r_1 = 0.9$ and $D_2/D_1=2$, Table 1). Species 2, initially homogeneously mixed with species 1, outcompetes species 1 once it reaches the leading edge: its faster dispersal enables it to reach the low-density edge where it can take advantage of the resources available, despite a disadvantage in growth rate. Once species 2 dominates the edge, species 1 is left behind in the high-density region where growth has stopped. Over time, the global frequency of species 1—blocked by species 2 from reaching the edge and incapable of growing further—decreases whereas species 2 frequency keeps increasing thanks to its edge domination (Fig. \ref{fig:fig1}C, left). These simulation results are consistent with experimental tests conducted here (Fig. \ref{fig:fig1}C, right) and also with the original experiment that led to evolution of hyperswarmers (\citealt{Ditmarsch2013}), which clearly showed that \textit{fleN} mutants would outcompete the wild-type to extinction given sufficient competition time on swarming plates.

According to the condition for success (Eq. 4) the evolutionary outcome is entirely determined from the growth and dispersal rates. Importantly, and similar to the expansion rate obtained for a monospecies traveling wave, the evolutionary outcome is independent of the carrying capacity of each species (See Fig. \ref{fig:figS4} for confirmation with numerical simulations). 

\begin{figure}[h!]
	\includegraphics[width=12cm]{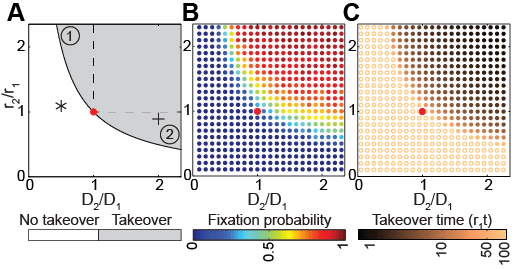}
	\caption{Faster-dispersing species take over expanding populations despite having a slower growth as long as $v_2>v_1$. A: Phase diagram of evolutionary outcome in $r_2/r_1$,$D_2/D_1$  space, with 2 subdomains of interest in the success domain: (1) is the domain of success with higher growth rate and slower dispersal. (2) is the domain of success with faster dispersal but lower growth rate. The (+) symbol represents the hyperswarmer phenotype with respect to wild-type phenotype, as measured in the experimental conditions (Table 1). The (*) symbol represents the wild-type phenotype with respect to the hyperswarmer phenotype. B: Fixation probability obtained from the stochastic model with death rate (stochastic simulations performed with $S=1$, $K=100$, $L=2\lambda_1$). C: Takeover time obtained from the deterministic model (simulations performed with $L=2\lambda_1$ and $S=0.2$). The empty orange circles represent the conditions where species 2 fails to takeover within the duration of the simulation ($r_1 t = 180$). In A-C, the red dot depicts the reference point ($D_2 = D_1$ and $r_2 = r_1$).}
	\label{fig:fig2}
\end{figure}

The success condition leads to a diagram that delineates a growth-dispersal space (Fig. \ref{fig:fig2}A, where the condition is expressed as $r_2/r_1 > \frac{1}{D_2/D_1}$). This diagram shows two trivial domains: when both growth and dispersal of species 2 are lower ($D_2<D_1$ and $r_2<r_1$), species 2 cannot outcompete species 1 because $v_2$ is always lower than $v_1$. Numerical simulations illustrate that for very low values of $D_2$ and $r_2$ species 1 continues to expand and travel at constant expansion rate whereas species 2 spreads out and stalls (Video 4, bottom left panel, and Fig. \ref{fig:figS5}A). When growth and dispersal of species 2 are greater ($D_2>D_1$ and $r_2>r_1$), species 2 takes over because $v_2$ is always higher than $v_1$: species 2 grows rapidly, moves to the front where it reaches the active layer and outcompetes species 1 (Video 4, top right panel, and Fig. \ref{fig:figS5}B). 

The two domains where one trait is higher and the other is lower are less trivial, but arguably more relevant. Because of the trade-off between dispersal and growth commonly found in nature (\citealt{Chuang2016}), a higher growth does not necessarily yield to evolutionary success: If species 2 grows faster than species 1 but disperses much slower ($r_1<r_2$ and $D_2<D_1 r_1/r_2$), then species 2 does not outcompete species 1. In other words, species 2 cannot take over the edge if its growth rate is not high enough to compensate a loss in dispersal ($r_1<r_2<r_1 D_1/D_2$). Domain 1 of Fig. \ref{fig:fig2}A shows that takeover occurs when species 2 disperses slower than species 1 only if its growth rate is sufficiently higher ($r_2>r_1 D_1/D_2$, Video 4, top left panel, and Fig. \ref{fig:figS5}C). Conversely, a slower growth does not mean takeover is impossible. If species 2 grows slower than species 1 but disperses sufficiently faster ($r_2<r_1$ and $D_2>D_1 r_1/r_2$), then it outcompetes species 1. In other words, there is takeover by species 2 if its growth rate is not too low ($r_1 D_1/D_2 < r_2 < r_1$) so the gain in dispersal can compensate the loss in growth. (Video 4, bottom left panel, and Fig. \ref{fig:figS5}D). Note that when species 2 replaces species 1 at the edge, the slope of the front changes accordingly (the length scale ranges from $\lambda_1 = \sqrt{D_1/r_1}$ to $\lambda_2 = \sqrt{D_2/r_2}$, see Fig. \ref{fig:figS6}).

In the model, the winning species takes over the front and replace the ancestor in the advancing front. However, the core of the population is not affected by the replacement that occurred at the edge. This results in the coexistence of two populations: the ancestor that remains in the initial spatial range and the competitor that occupies the newly extended range.

The simple condition for success, $v_2>v_1$, makes quantitative predictions of evolutionary outcome; those predictions hold true in our experimental system despite intricacies such as the large-scale branching and small-scale turbulence. Hyperswarmers have a $\sim 100\%$ increase in dispersal ($D_2/D_1 \sim 2$) that comes at a $\sim 10\%$ growth rate cost ($r_2/r_1=0.9$) (Video 2 and Table 1). Therefore, the experimental system falls into domain 2 of the evolutionary outcome diagram (cross symbol in Fig. \ref{fig:fig2}A).

We measured the frequency of hyperswarmers within the first millimeter of the colony from video frames; it increased exponentially with a rate of $0.39 \pm 0.08 \text{h}^{-1}$ (SD), which is in quantitative agreement with the theoretical expression of the relative fitness (Eq. 3 and Fig. \ref{fig:figS7}). Hyperswarmers introduced into an expanding wild-type colony spread within a wild-type branch (Fig. \ref{fig:figS8}), reach the tip of the branch, and take over the population (Video 5A) resembling our simulations (Fig. \ref{fig:figS5}D).

Hyperswarmers evolved from a wild-type swarming colony (\citealt{Ditmarsch2013}) and could take over the ancestral population thanks to a greater dispersal. To test our model, we asked whether this process was reversible: Could wild-type cells dominate the edge of an expanding hyperswarmer colony thanks to their greater growth rate? Our model predicted that wild-type cells would be unable to take over the hyperswarmer population edge since in this case $v_2<v_1$ ($D_2/D_1=0.5$ and $r_2/r_1=1.1$, see star symbol in Fig. \ref{fig:fig2}A). This was confirmed experimentally: wild-type cells introduced in a hyperswarmer colony simply spread out and were rapidly outpaced at the edge by the hyperswarmers (Video 5B). Note that if $v_2<v_1$, species 2 cannot replace the species 1, not even by forming a block at the front. According to the model some individuals of species 1 will diffuse through and reach the edge to eventually take over (See simulations of this process in Fig. \ref{fig:figS9}).

\subsection*{Success rule valid despite phenotypic variability}
Even in mono-species systems, individuals with identical and defined genotypes can still display phenotypic variation, such as a varying number of flagella (\citealt{Deforet2014, Waite2016}). To study whether phenotypic variation had an effect on evolutionary outcome, we introduced non-heritable fluctuations in birth and death events as well as in the dispersal processes. These phenotypic variations were modeled as stochastic distributions around the mean population value, which is determined by the strain’s genotype (see Appendix C: Stochastic Modeling).

Our simulation results suggest that the success rule, $v_2>v_1$, despite having been derived from deterministic assumptions, holds even in stochastic situations. The transition at $v_2=v_1$ was, however, more gradual (Fig. \ref{fig:fig2}B, Fig. \ref{fig:figS10}, Fig. \ref{fig:stochastic}): the zone of transition broadened as stochasticity increased because, as expected from other stochastic studies (\citealt{Otto1997, Gillespie2010}), stochasticity allowed for a non-zero probability of deleterious mutants ($v_2<v_1$) to take over and beneficial mutants ($v_2>v_1$) had a non-zero probability of failing to take over. Larger carrying capacities lessened the stochastic effects and sharpened the transition zone, again as expected from previous stochastic analyses (\citealt{Otto1997, Gillespie2010}). Importantly, however, the rule $v_2>v_1$ could still predict takeover of the population edge even with different carrying capacities (Fig. \ref{fig:stochastic}C).

We confirmed the generality of the success rule further by carrying out evolutionary simulations where mutations randomly arise at division. We considered two schemes: i) mutations that change growth and/or dispersal relative to the ancestor phenotype but do so in an uncorrelated way; ii) mutations that change growth and dispersal considering that the two traits are linearly correlated (linear trade-off) but independent of the ancestor phenotype. In the case of uncorrelated mutations, populations evolved—on average—towards a greater expansion rate $v=2\sqrt{rD}$ (Fig. \ref{fig:simuevo}C). When the two traits were constrained by a trade-off, evolution converged to the value along the trade-off line that maximized the expansion rate $v=2\sqrt{rD}$ (Fig. \ref{fig:simuevo}D-G). In summary, the several types of stochastic simulations conducted all supported that evolution of the population edge obeys the rule $v_2>v_1$.

\subsection*{The role of spatial structure and founder effect}
We then investigated whether our model would account for other factors that could affect competition in biologically relevant scenarios. For example, in most evolutionary scenarios where competition starts with a mutation, the size of the mutant population is initially very low (1 individual) whereas a competing species introduced by external processes (e.g. human intervention) can start at higher densities. Also, the initial location of the mutant species matters because resources are not evenly distributed in nature, and a mutant species may take over faster if it is introduced in the resource-rich leading edge than if it is introduced in deprived regions where it will take longer to grow to domination. In summary, species 2 should take longer to take over (i) when it is introduced further from the edge where resources are already limited or (ii) when its initial size is small. Our model sets the conditions for whether species 2 can successfully take over (Eq. 4, Fig. \ref{fig:fig2}A) but does not give us the time necessary for establishing at the edge. 

To investigate how the time to takeover depends on the location and initial size of the introduced population, we modeled the introduction of species 2 into a traveling wave formed by species 1. We assumed an initial density $S$  across a small interval at a distance $L$ from the edge for species 2 (Fig. \label{fig:figB1}), and we determined the time needed to outnumber (full sweep) species 1 at the front. Numerical simulations revealed that the general rule, $v_2>v_1$, holds for all initial conditions given sufficient time (Fig. \ref{fig:fig2}C). The time required, however, depends on the initial conditions, increasing approximately linearly with the distance $L$ from the front and decreasing sub-linearly with the initial density $S$ (Fig. \ref{fig:phase} and \ref{fig:delay}).

To better distinguish the factors that influence the time required for takeover we considered two steps: first, we considered that species 2 disperses until it reaches the active layer. The time for species 2 to reach the edge depends on the distance from the introduction point to the front—a distance that increases constantly because species 1 is itself advancing—and also on the initial width of species 2. Second, once species 2 reaches the active layer it must grow to outnumber species 1. When the introduction is sufficiently far from the edge the time of takeover, $t_t$, is:
\begin{equation}
t_t \sim \alpha L - \beta \log(S)
\end{equation}
where $\alpha$ and $\beta$ depend on the parameters $D_1$, $D_2$, $r_1$ and $r_2$ (see Appendix B: Approximate predictions). This analysis confirmed simulation results that the time to takeover depends linearly on $L$ but only sub-linearly on $S$ (Fig. \ref{fig:phase} and \ref{fig:delay}), highlighting that the distance to the edge is key to evolutionary success. 

\subsection*{Experimental validation}
\begin{figure}[h!tp]
	\includegraphics[width=8cm]{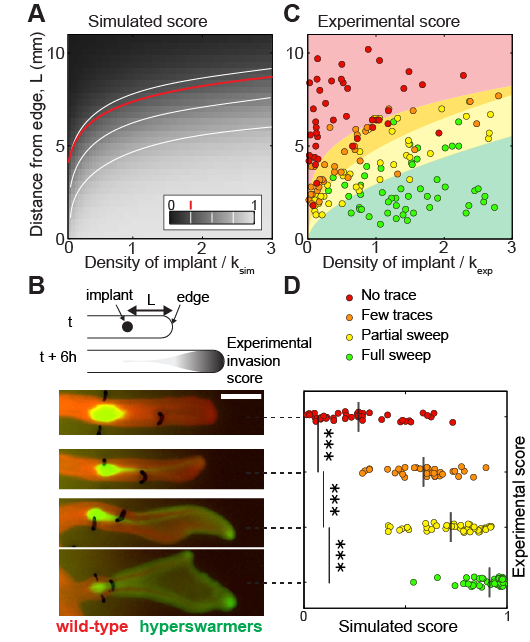}
	\caption{Likelihood of fixation increases with the initial size of the introduced population and its proximity to the population edge. A: Simulation results of introducing species 2 into an expanding species 1. The color scale represents the simulated sweep score, i.e. the frequency $f$ of the introduced species (species 2 with $D_2=2D_1$ and $r_2=0.9r_1$) at the population edge $T=6\text{ h}$ after introduction. White lines are iso-frequency lines for $f = 0.25$, $0.5$, and $0.75$. The red line is from Eq. 6 for $f = 0.25$. B: Laboratory experiments where the hyperswarmer (in green) was introduced at varying initial densities and distances to the edge of an expanding swarm of wild-type \textit{P. aeruginosa} (in red). Scale bar is 5 mm. Leftmost marks depict the location of the hyperswarmer introduction; rightmost marks locate the position of the front of the \textit{P. aeruginosa} population at the time of hyperswarmer implantation. The four snapshots represent four experimental replicates. C: Experimental sweep success evaluated visually at 6 hours after hyperswarmer introduction. Background colors represent results from multinomial logistic regression (see details in Appendix D: Statistical analysis). In agreement with the theory, sweep success is lower for large distances from the front and smaller initial densities. D: Comparison of simulated and experimental sweep scores for each experimental replicate. The grey vertical lines represent the average simulated sweep score and the p-values are $< 10^{-3}$ (Kruskal-Wallis test).}
	\label{fig:fig3}
\end{figure}

We then tested these findings in our experimental system. We manipulated the distance to the edge ($L$) and the density ($S$) of a small population of hyperswarmers introduced into an expanding wild-type population, and we compared the experimental results to the corresponding simulations. In simulations, the evolutionary outcome was calculated as the frequency $f$ of species 2 at the edge of the population 6 hours after implantation (Fig. \ref{fig:fig3}A). In the experiments, we ranked the evolutionary outcome after 6 h of expansion as no trace, few traces, partial sweep and full sweep according to the amount of hyperswarmers visible at the edge (Fig. \ref{fig:fig3}B, see details in Methods and Models section). The experiments confirmed the dominant role of $L$ compared to $S$ in determining the time of takeover (Eq. 5), which is evident from the concave shape of the evolutionary scores (Fig. \ref{fig:fig3}C, see Appendix D: Statistical analysis).

The intuition behind the concave shape is that when the initial distance from the edge is too long then the mutant may not be able to take over within biologically relevant time, even if its initial size is large. The shape of the iso-frequency contour lines can be calculated from the simplified two-step model of takeover described above and is given by 
\begin{equation}
S \sim e^{L/L_0}
\end{equation}
where $L_0 = 4 D_2 v_1/(3 v_1^2 + v_2^2)$ is the characteristic length of these lines. In the case of \textit{P. aeruginosa} and its hyperswarmers, $L_0 = 1.1 \pm 0.07 \text{ mm}$ (SD). The results from our hyperswarmer experiments agree with the theoretical model (Fig. \ref{fig:fig3}A and C; compare to lines of constant mutant frequency), indicating that the two-species F-KPP model, in spite of its simplifying assumptions and despite any intricacies of the experimental system (e.g. the swarming population is tri-dimensional; bacterial cells tend to lose motility as they lose access to resources inside the population, which freezes the spatial organization), is sufficiently general to describe evolutionary dynamics in swarming colonies.

\section*{Discussion}
We showed that the multiple-species extension of the F-KPP equation (\citealt{Lewis2002, Okubo1989, Pigolotti2013}) produces a simple mathematical rule that predicts the evolutionary outcome in an expanding edge depending on the growth and dispersal rates of the competing species. The problem of evolution in an expanding population have been investigated before both theoretically (e.g \citet{Burton2010, Phillips2015} and empirically (e.g. \citet{Phillips2006}), but its simple solution, the inequality $v_2>v_1$, had not been—to the best of our knowledge—presented this way before.

Our model relies on the assumption that per-capita growth rate is maximal at the edge, where the population density is the lowest, and that dispersal ability is independent of population density. This assumption is valid within the first centimeters of our experimental swarming colony, where bacteria are quite motile and active. However, deep inside the colony, various processes at play hinder a quantitative analysis of population dynamics: the colony can progress from swarming colony into a biofilm-like mode that greatly lowers dispersal, starved bacteria secrete molecules that are auto-fluorescent, and the long-term maturation kinetics of the fluorescent proteins used (GFP and DsRed) can vary. Therefore, the situation may start to differ from the idealized model. Once the edge has passed the dynamics can be quite different: in the resource-depleted region, the population is denser and covers the entire available area (Fig. \ref{fig:figS2}). The spatial structure and dispersal are less relevant and the evolutionary fate of a new mutant is determined by high-density dynamics. The faster growing wild-type can catch up or new mutants carrying compensatory mutations that thrive in low resource environments may even appear (\citealt{Yan2017}).  These successional dynamics may be a natural product of evolution, so long as there is someplace for the faster disperser to utilize.
 
\begin{figure}[h!tp]
	\includegraphics[width=8cm]{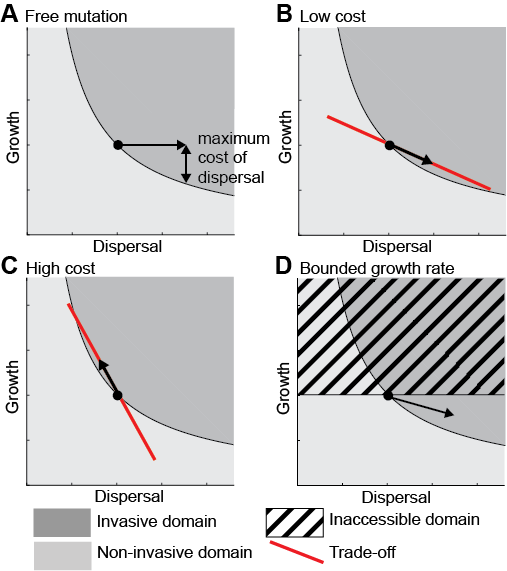}
	\caption{Evolutionary predictions from success rule in four scenarios. A: Illustration of the maximal cost of an increased dispersal. B and C: Linear trade-off between growth and dispersal that corresponds to a low cost of dispersal (B) and high cost of dispersal (C). D: The growth rate is bounded by physiology.}
	\label{fig:fig4}
\end{figure}

Our model makes a key conclusion: the outcome at the edge of expanding population can be independent of the system’s carrying capacity, because there the competition dynamics rely on the low-density of the population at the expanding edge. The success rule $v_2>v_1$ allows determination of the maximal cost in growth $|\Delta r|_\text{max}$ that a mutant can afford to pay for faster dispersal and still be able to dominate the edge of the expanding population (Fig. \ref{fig:fig4}A):
\begin{equation}
|\Delta r|_\text{max} = r_1 \Big( 1-\frac{D_1}{D_1 + \Delta D}\Big)
\end{equation}
where $\Delta D$ is the difference between dispersal rate of the mutant and its ancestor. Eq. 7 quantifies exactly what the trade-off between $r$ and $D$ would need to be in order to evolve greater dispersal at the front. While this is not something that is easily confronted with data it is worth noting that there must be general mechanisms to sustain this in populations that have exhibited such evolutionary increases in dispersal. The evolutionary experiment that originally created the hyperswarmers always produced single point mutations in \textit{fleN}, a gene that regulates flagella synthesis, and all had slower growth that the wild type (\citealt{Ditmarsch2013}). We never observed mutants that evolved faster dispersal without a growth cost, even though we repeated the experiment dozens of times. Perhaps other mutants could increase dispersal even more, but were not favored because they either carried costs higher than $|\Delta r|_\text{max}$ or because they required evolution through more mutational steps.

Hyperswarmer mutants paid a growth cost for synthesizing and operating their multiple flagella but—without affecting their competitive ability—dispersed faster than wild-type bacteria (domain 2 of Fig. \ref{fig:fig2}A). It seems plausible that the supplemental flagella are functional, and their operation adds a cost, but we cannot provide evidence on this point. Extensive work showed that the mutation in \textit{fleN} increases the number of flagella and slows down the growth rate (\citealt{Ditmarsch2013}), but that growth cost could be due to the burden of synthesizing extra flagella or to the extra energetic burden of their operation. Untangling the two remains an interesting problem, but solving it requires molecular biology work beyond the scope of this paper. 

The success rule $v_2>v_1$ is the adaptive function that could be combined with a fitness set (\citealt{Mazancourt2004}) to predict the co-evolution of growth and dispersal in expanding populations. For example, if the quantitative knowledge on the molecular, cellular, or physiological mechanisms of a trade-off between growth and dispersal can be represented by a line in $(r,D)$ space the slope of that line, $dr/dD$, represents the cost of dispersal. When the trade-off is subtle, the slope is shallow and we predict that the population will evolve to disperse faster with a lower growth rate (Fig. \ref{fig:fig4}B). Conversely, when the trade-off is strong, the slope is steep and we predict that the population will evolve a higher growth rate and slower dispersal (Fig. \ref{fig:fig4}C). According to this model, the \textit{P. aeruginosa} system has a subtle trade-off: the improved dispersal advantage of hyperswarmers is $\sim 100\%$ but costs only $\sim 10\%$ of their growth rate relatively to wild-type (\citealt{Ditmarsch2013}).

Nature abounds with examples of trade-off between growth and dispersal (\citealt{Chuang2016}). Evolution in expanding populations often selects for better dispersal and slower growth (see examples cited in the introduction). Our model predicts that a species with faster growth but slower dispersal should be able to take over (domain 1 of Fig. \ref{fig:fig2}A), but we never observed these cases in our experimental system. And, beyond observations by another group in laboratory experiments with \textit{Escherichia coli} (\citealt{Fraebel2017}), we could not find examples in nature either. The reason for not finding evolution of rapid growers that disperse slower may be population history: Empirical and theoretical studies of range expansions suggest that only dispersal can be improved in expanding populations (\citealt{Burton2010, Hallatschek2014, Perkins2013, Travis2002}); invasion of new niches is possibly a rare event, whereas competition within a confined, but relatively homogenous environment is more common. In such situations, selection is not on dispersal but on growth, which means that—in most species—growth rates may already be close to their physiological maximum (Fig. \ref{fig:fig4}D). Individuals challenged to overcome spatial structure may only have dispersal-related traits left to improve. Moreover, while the trade-off between growth and dispersal may be found and seem logical, a comparative analysis of dispersal in terrestrial and semi-terrestrial animals suggested that dispersal and fecundity may be positively correlated (\citealt{Stevens2014}).

Margins of an expansion front, with low population density and strong density gradient, are prone to dramatic evolutionary processes such as spatial sorting and expansion load, which can be generalized using the rule $v_2>v_1$:
(i) A mutant with a higher $v$ will take over the population margin. This is a generalisation of the spatial sorting effect, where better dispersers ($D_2>D_1$, with $r_2=r_1$) accumulate at the population margins (\citealt{Shine2011}). 
(ii) Genetic drift in marginal populations can promote accumulation of deleterious mutations in the form of an expansion load (\citealt{Hallatschek2008, Peischl2013}). In our framework, this corresponds to the stochastic case where a mutant takes over the margin with a lower growth rate $r_2<r_1$ and $D_2=D_1$). We demonstrated that stochasticity can allow, more generally, a mutant with lower $v$ to take over.
This model suggests that the fitness (net balance between growth and death) can be replaced with a traveling wave fitness $v$, which combines growth rate and dispersal rate. An increase in $v$ leads to spatial sorting, and stochastic effects can lead to accumulation of low $v$ mutants at the edge (expansion load).

Our results produced a general and simple relationship that determines the maximum growth cost allowed for faster dispersal. This appealing simple rule is bound to our model assumptions, which apply more directly to microbial systems. Future work should address how more complex biological systems deviate, or not, from those assumptions. Possible expansions include situations where the growth rate is not maximal at the edge (\citealt{Korolev2015, Perkins2013}), or where uncertainty about the quality of resources beyond the edge front factor in. The only interaction we considered here lies in the shared carrying capacity. Extension to more complex systems could also include explicit interactions between individuals, such as in competition-colonization models, introduced to address the question of coexistence in spatially structured environments (\citealt{Tilman1994}).

In conclusion, our study provides theory to determine the evolutionary outcome of competition in an expanding population, which can be extended with trade-off constraints observed for each particular system. Every model requires simplifying assumptions, and ours is certainly not an exception. In systems that respect those assumptions, the success rule could be used to predict evolution in expanding populations. Systems in this category may include the growth of cancer tumors and invasion of non-native species in ecosystems.

\section*{Acknowledgments}

The authors are grateful to Florence Débarre for her valuable comments on the manuscript.
This work was supported by National Science Foundation award MCB-1517002/NSF 13-520 (to J.B.X) and by National Institutes of Health Grant R00CA191021 (to C.C.-F.). The funders had no role in study design, data collection and analysis, decision to publish, or preparation of the manuscript.

\newpage{}

\renewcommand{\theequation}{A\arabic{equation}}
\renewcommand{\thetable}{A\arabic{table}}
\renewcommand{\thefigure}{A\arabic{figure}}
\setcounter{equation}{0}  
\setcounter{figure}{0}
\setcounter{table}{0}

\section*{Appendix A: Analytical solution for the condition of success}
\subsection*{Formulation of the mathematical model}

\begin{figure}[htp]
	\centering
	\includegraphics[width=10cm]{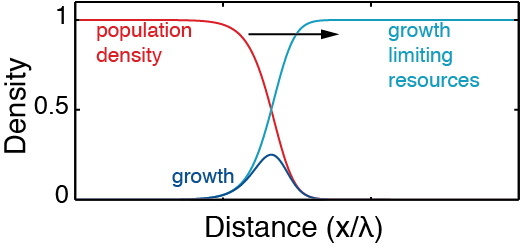}
	\caption{Diagram of the traveling wave solution of the F-KPP equation.}
	\label{fig:figA1}
\end{figure}

The traveling wave solution of the F-KPP equation is driven by growth 
at the edge of the population range. In the figure \ref{fig:figA1}, the population density $u$ is represented in red, the growth limiting resources (proxied by the difference between the carrying capacity and the population density) $1-u$ in light blue, the growth $ru(1-u)$ in dark blue. The black 
arrow depicts the direction of expansion.

We model the density of the two species (densities $u_1$ and $u_2$) by two coupled F-KPP 
equations with different growth rates ($r_1$ and $r_2$) and different diffusion 
rates ($D_1$ and $D_2$). The two species interact only by competing for common 
resources.
\begin{equation}\label{eq:first}
\begin{aligned}
\frac{\partial u_1}{\partial t} & = r_1 u_1(1-u_1-u_2) + D_1 \frac{\partial^2 
	u_1}{\partial x^2}\\
\frac{\partial u_2}{\partial t} & = r_2 u_2(1-u_1-u_2) + D_2 \frac{\partial^2 
	u_2}{\partial x^2}
\end{aligned}
\end{equation}
We aim to define the range of parameters ($D_1$, $D_2$, $r_1$, $r_2$) that 
allows the takeover of an established traveling wave (species 1) by another 
species (species 2).

\subsection*{Eigenvalue problem}
We reproduce the same arguments as in \citet{Korolev2015} but with $r_1 \neq 
r_2$. We introduce the functions $u(t,x) = 
u_1(t,x) + u_2(t,x)$, $g_1(t,x)=r_1 (1-u(t,x))$ and $g_2(t,x)=r_2 (1-u(t,x))$.
\begin{equation}\label{eq:g}
\begin{aligned}
\frac{\partial u_1}{\partial t} & = g_1(u) u_1 + D_1 \frac{\partial^2 
	u_1}{\partial x^2}\\
\frac{\partial u_2}{\partial t} & = g_2(u) u_2 + D_2 \frac{\partial^2 
	u_2}{\partial x^2}
\end{aligned}
\end{equation}

To find the condition of takeover, we search for the condition of 
divergence of the fraction of secondary species:
\begin{equation}
f(t,x) = \frac{u_2(t,x)}{u_1(t,x) + u_2(t,x)}
\end{equation}

\citet{Korolev2015} assumes $f\ll1$ and demonstrates that the condition of 
takeover can be found by solving an eigenvalue problem. Following the exact 
same steps, we find that in the moving 
reference frame (where the space variable is $\xi=x-v_1 t$) traveling at the 
velocity $v_1=2\sqrt{r_1 D_1}$ the eigenvalue problem is the following:
\begin{equation}\label{eq:eig1}
\tilde r f = D_2 f'' + (v_1 + 2 D_2 \frac{c'}{c})f' + (D_2-D_1)\frac{c''}{c}f + 
(g_2-g_1)f
\end{equation}
where primes denote the derivative with respect to $\xi$, and $\tilde r$ is an 
eigenvalue.
After another change of variables established by Korolev, equation 
\ref{eq:eig1} becomes
\begin{equation}\label{eq:eig2}
\tilde r  \psi = D_2 \psi'' + \Big(g_2(c) - \frac{v_1^2}{4D_2}\Big)\psi
\end{equation}

\subsection*{Condition of takeover}
Following Korolev's reasoning (\citealt{Korolev2015}), we find that the largest eigenvalue is found at 
large $\xi$, and its value is given by the maximal value of $g_2(c)$:
\begin{equation}\label{eq:eigval}
\begin{aligned}
\tilde r _{\text{max}} & = r_2 - \frac{v_1^2}{4D_2}\\
& = \frac{v_2^2-v_1^2}{4D_2}
\end{aligned}
\end{equation}
If $\tilde r _{\text{max}}<0$ then the secondary species will never take over the primary species.
If $\tilde r _{\text{max}}>0$ then the secondary species will eventually take over the primary species.
Therefore, the condition of takeover corresponds to
\begin{equation}\label{eq:condition}
v_2 > v_1 
\end{equation}
with $v_1 = 2 \sqrt{r_1 D_1}$ and $v_2 = 2 \sqrt{r_2 D_2}$.

\newpage{}
\renewcommand{\theequation}{B\arabic{equation}}
\renewcommand{\thetable}{B\arabic{table}}
\renewcommand{\thefigure}{B\arabic{figure}}
\setcounter{equation}{0}  
\setcounter{table}{0}
\setcounter{figure}{0}

\section*{Appendix B: Approximate predictions}
The equation \ref{eq:condition} gives the condition of takeover, but not the 
time of takeover, which depends on the
size $S$ of the introduced population and the location of the introduction (distance $L$ from 
the front). 

\subsection*{Definitions of $S$ and $L$}
We first consider a traveling wave formed by the species 1. Then the species 2 
is introduced at a distance $L$ from the population front, as depicted in the figure below.

\begin{figure}[h!]
	\centering
	\includegraphics[width=8cm]{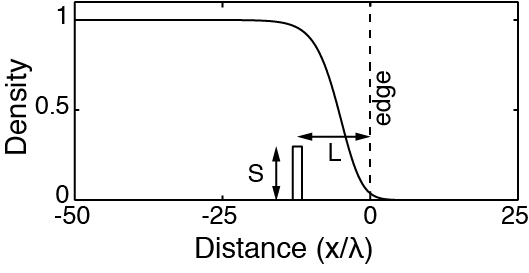}
	\caption{Diagram of the invasion scheme: the species 2 is implanted at a distance $L$ from the edge, with a density $S$.}
	\label{fig:figB1}
\end{figure}

The front position is defined as the location where $u_1$ is equal to 5\% of the carrying 
capacity, without loss of generality.
This threshold is arbitrarily selected but it is not essential for later 
conclusions.
The position of the front at the time of introduction can be taken as the origin 
of the x-axis,
and the time of the introduction sets the origin of time (i.e. $u_1(0,0)=0.05$).
The introduction imposes a density $u_2(x,0)=S$ for 
$-L-\lambda_1<x<-L+\lambda_1$, and $u_2(x,0)$ everywhere else.
Then we consider the F-KPP processes for both species simultaneously.
We set $r_1=1$ and  $D_1=1$ by rescaling space ($x \rightarrow x/\lambda_1$) and 
time ($t \rightarrow r_1 t$): the remaining parameters are $D_2$ and $r_2$.
The time $t_i$ at which the species 2 takes over the species 1 at 
the front is defined by:
\begin{equation}
u_2(x_f,t_i)>u_1(x_f,t_i)
\end{equation}
with the position of the front defined as 
\begin{equation}
u_1(x_f,t)+u_2(x_f,t)=0.05
\end{equation}

An approximate time of takeover can be calculated by considering two consecutive steps. First, 
the introduced individuals of species 2 spread through spatial diffusion until 
the 
density $f$ of species 2 at the edge reaches a maximum. During this step we 
assume that species 2 does not grow.
Then, the individuals of species 2 that reached the edge follow the traveling 
wave. They have access to the resources and therefore they proliferate until 
they take over the species 1.

\subsection*{Diffusive spreading }
The introduction is initially spatially limited, and performed at $x=0$, at 
distance 
$L$ from the edge. Then species 2 diffuses out, and its density is:
\begin{equation}
u_2(x,t) = N^0_2 \frac{e^{-\frac{x^2}{4D_2 t}}}{\sqrt{4\pi D_2 t}}
\end{equation}
where $N^0_2$ is the total size of the introduced population. To compare the theoretical and 
experimental results, we rescale the total size of the introduced population using its 
spatial extension: 
\begin{equation}\label{eq:n0}
N^0_2=2 \lambda_1 S
\end{equation}

The density of species 2 at the edge of the traveling wave formed by species 1 
is $u_2(x=v_1 t+L, t)$ and it reaches a maximum at 
\begin{equation}\label{eq:tdiff}
t = t_\text{diffusion} \equiv \frac{\sqrt{D_2^2+v_1^2 L^2}-D_2}{v_1^2}
\end{equation}

At this time, the density of species 2 at the edge is
\begin{equation}\label{eq:n2max}
u_{2,\text{diffusion}} = 2 \lambda_1 S \frac{e^{-\frac{(v_1 
			t_\text{diffusion}+L)^2}{4D_2 t_\text{diff}}}}{\sqrt{4\pi D_2 t_\text{diff}}}
\end{equation}

Since the edge of the population range is defined as the location where the 
total density is 5\% of the carrying capacity,
the fraction of species 2 at the edge is 
\begin{equation}\label{eq:fmax}
f_{\text{diffusion}} = \frac{2 \lambda_1 S}{0.05} \frac{e^{-\frac{(v_1 
			t_\text{diffusion}+L)^2}{4D_2 t_\text{diffusion}}}}{\sqrt{4\pi D_2 
		t_\text{diffusion}}}
\end{equation}

\subsection*{Growth at the edge }
At the edge, the densities of species 1 and species 2 are much 
smaller than the carrying capacity, so we consider that they grow exponentially.
If species 1 and 2 grow exponentially with growth rates $r_1$ and $r_2$, 
respectively, then the evolution of the frequency $f$ is given by the logistic 
equation with a growth rate $r_2 - r_1$.
We study the early times of takeover, hence $f$ remains much smaller than 1. 
Therefore, we simply write:
\begin{equation}\label{eq:r1r2}
\frac{\partial f}{\partial t} = (r_2-r_1)f
\end{equation}
Here, we study the dynamics of the frequency $f$ in the moving reference frame 
(see Appendix A) at short times, so we can assume that 
$f$ follows the dynamics given by the 
eigenvalue equation \ref{eq:eig1} and therefore grows at a rate corresponding 
to the maximal eigenvalue given in equation \ref{eq:eigval}. Then we can 
replace $r_2-r_1$ with $\tilde r _\text{max}$ in equation \ref{eq:r1r2}.

The dynamics of $f$ is therefore:
\begin{equation}\label{eq:ft}
f(t) =f_{\text{diffusion}} e^{\tilde r _\text{max} t}
\end{equation}

\subsection*{Frequency of species 2 at the edge}
If the introduction is performed far enough from the edge ($L>D_2/v_1$), then 
equation 
\ref{eq:tdiff} becomes simply $t_\text{diffusion} \simeq L/v_1$ and equation 
\ref{eq:fmax} 
becomes:
\begin{equation}\label{eq:fdiff_simple}
\begin{aligned}
f_{\text{diffusion}} & \simeq \frac{2 \lambda_1 S}{0.05} \frac{e^{-\frac{L 
			v_1}{D_2}}}{\sqrt{4\pi D_2  
		L/v_1}} \\
& \simeq \frac{S}{0.05}\frac{1}{\sqrt{\pi}} 
\sqrt{\frac{D_1}{D_2}}\sqrt{\frac{v_1}{r_2 L}} e^{-\frac{L v_1}{D_2}}
\end{aligned}
\end{equation}

Since the $\sqrt{L}$ evolves much more slowly then $e^{-\frac{L v_1}{D_2}}$, we 
consider that $L$ is constant in this term (we use $L=\tilde L = 5 \text{mm}$ 
for the 
experimental validation):
\begin{equation}\label{eq:fdiff_simple2}
f_{\text{diffusion}}  \simeq \frac{S}{0.05}\frac{1}{\sqrt{\pi}} 
\sqrt{\frac{D_1}{D_2}}\sqrt{\frac{v_1}{r_2 \tilde L}} e^{-\frac{L v_1}{D_2}}
\end{equation}

The frequency of species 2 after a time $t$ is given by equation \ref{eq:ft} 
but 
using the shifted time $t-t_{\text{diffusion}}$ that accounts for the time 
required for species 2 to reach the edge.
\begin{equation}\label{eq:ft2}
\begin{aligned}
f(t) & \simeq \frac{S}{0.05}\frac{1}{\sqrt{\pi}} 
\sqrt{\frac{D_1}{D_2}}\sqrt{\frac{v_1}{r_2 \tilde L}} e^{-\frac{L v_1}{D_2}} 
e^{\tilde r _\text{max} (t-t_{\text{diffusion}})} \\
& \simeq \frac{S}{0.05}\frac{1}{\sqrt{\pi}} 
\sqrt{\frac{D_1}{D_2}}\sqrt{\frac{v_1}{r_2 \tilde L}} 
e^{\frac{v_2^2-v_1^2}{4D_2} t} e^{-\frac{3v_1^2+v_2^2}{4D_2 v_1}L}  \\
\end{aligned}
\end{equation}
Hence the iso-frequency lines in Fig. \ref{fig:fig3}A are given by $S \sim e^{L/L_0}$, 
with 
$L_0=\frac{4D_2 v_1}{3v_1^2+v_2^2}$ (equation 5 of the main text).

The iso-frequency for $f=0.25$ is plotted in the figure below 
(red dashed line), 
together with the numerical results. The disagreement with the numerical 
results 
is mostly due to the simplicity of the model: first the species 2 
diffuses out and reaches a maximum density at the edge, then it grows at the 
edge. In reality, species 2 starts growing as soon as it reaches the edge and gains access to resources.
To account for this neglected growth term, we propose a simple correction:
we replace the frequency at the edge $f_{\text{diffusion}}$ with $f_{\text{diffusion}}(1+ 
t_{\text{diffusion}}r_2/2)$ in equation \ref{eq:ft}.
The term  $f_{\text{diffusion}}t_{\text{diffusion}}r_2/2$ comes from the 
approximate integration of the growth of the frequency $f$ at the edge from 
$t=0$ ($f=0$) to $t=t_\text{diffusion}$ ($f=f_\text{diffusion}$).
Including this growth term improves the agreement with the numerical simulations 
(red solid line of the figure below).

\begin{figure}[h!]
	\centering
	\includegraphics[width=6cm]{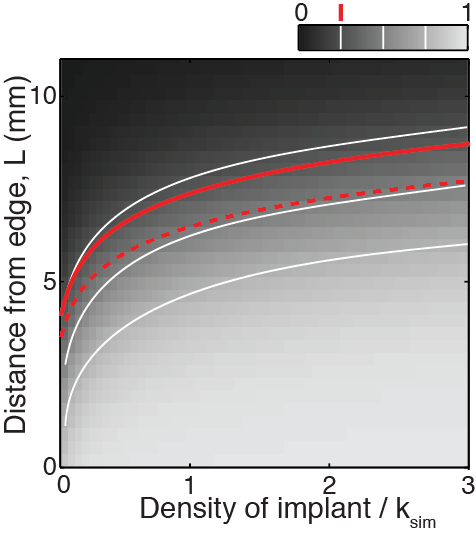}
	\caption{Comparison between numerical and theoretical results.
		The color scale represents the simulated score (the ratio $f$ of 
		species 2 density over total population density) at the front of the expansion range, at $t=6$ hours after implantation. The secondary species is characterized by $D_2=2D_1$ and 
		$r_2=0.9r_1$. The white lines are iso-frequency contour lines for simulated 
		score $f = 0.25$, 0.5, and 0.75. The red dashed line is the $f=0.25$ contour line from equation \ref{eq:ft2}. The red solid line is the $f=0.25$ contour line corrected for growth.}
	\label{fig:numerics}
\end{figure}

\subsection*{Time of takeover}
If $L>D_2/v_1$, the time of diffusion from implantation to the edge is
\begin{equation}
t_\text{diffusion}\simeq L/v_1
\end{equation}

The time of growth $t_\text{growth}$ to reach a certain frequency $f$ is:
\begin{equation}\label{eq:timeG}
\begin{aligned}
t_\text{growth} & \simeq  \frac{1}{\tilde r 
	_\text{max}}\log\bigg(\frac{f}{f_{\text{diffusion}}}\bigg) \\
& \simeq \frac{1}{\tilde r _\text{max}}\bigg(\log(0.05f/S) + 
\frac{1}{2}\log(\frac{\pi r_2 \tilde{L} D_2}{D_1 v_1}) + \frac{Lv_1}{D_2}\bigg)
\end{aligned}
\end{equation}

Overall, the time of takeover $t_t = t_\text{diffusion} + t_\text{growth}$ 
scales as (equation 5 of the main text):
\begin{equation}
t_t \sim \alpha L  - \beta \log(S)
\end{equation}
with 

\begin{equation}
\begin{aligned}
\alpha & = \frac{4v_1}{v_2^2-v_1^2} \\
\beta & = \frac{4D_2}{v_2^2-v_1^2}
\end{aligned}
\end{equation}

\newpage{}
\renewcommand{\theequation}{C\arabic{equation}}
\renewcommand{\thetable}{C\arabic{table}}
\renewcommand{\thefigure}{C\arabic{figure}}
\setcounter{equation}{0}  
\setcounter{table}{0}
\setcounter{figure}{0}

\section*{Appendix C: Stochastic modeling}
The model presented above is deterministic. We introduce stochasticity by updating species counts $N_1$ and $N_2$ using distribution functions based on equations \ref{eq:first}.
\subsection*{Logistic Growth} At each time step, the counts of species 1 and 2 are updated to account for logistic growth:
\begin{equation}\label{eq:LG}
\begin{aligned}
N_1(x,t+dt) &\rightarrow N_1(x,t) + \text{Poisson}(r_1 \mathop{dt} N_1(x,t) (K-N_1(x,t) - N_2(x,t))/K) \\
N_2(x,t+dt) &\rightarrow N_2(x,t) + \text{Poisson}(r_2 \mathop{dt} N_2(x,t) (K-N_2(x,t) - N_2(x,t))/K)
\end{aligned}
\end{equation}
where $\text{Poisson}(\lambda)$ is the Poisson distribution function, with parameter $\lambda$.

\subsection*{Exponential Growth + Death} To introduce a net death rate, we develop the expression of the deterministic
logistic growth into a (positive) exponential growth term and a (negative) density dependent death term.
\begin{equation}
\begin{aligned}
r_1 N_1.*(K-N_1-N_2)/K  &=  r_1 N_1  -  r_1/K N_1 (N_1 + N_2) \\
r_2 N_2.*(K-N_1-N_2)/K  &=  r_2 N_2  -  r_2/K N_2 (N_1 + N_2)
\end{aligned}
\end{equation}
Therefore, at each time step, species counts are updated in the following way:
\begin{equation}\label{eq:GD}
\begin{aligned}
N_1(x,t+\mathop{dt}) &\rightarrow N_1(x,t) + \text{Poisson}(r_1 \mathop{dt} N_1(x,t)) \\
N_1(x,t+\mathop{dt}) &\rightarrow N_1(x,t) - \text{Binomial}(N_1(x,t), r_1 \mathop{dt} /K (N_1(x,t) + N_2(x,t)) \\
N_2(x,t+\mathop{dt}) &\rightarrow N_2(x,t) + \text{Poisson}(\mathop{dt} r_2 N_2(x,t)) \\
N_2(x,t+\mathop{dt}) &\rightarrow N_2(x,t) - \text{Binomial}(N_2(x,t), r_2 \mathop{dt} /K (N_1(x,t) + N_2(x,t))
\end{aligned}
\end{equation}
where $\text{Binomial}(n,p)$ is the binomial distribution function, with parameters $n$ (number of trials) and $p$ (success probability in each trial).

\subsection*{Diffusion} At each time step, a random fraction of individuals of species 1 in deme $x$ are randomly moved into demes $x+dx$ and $x-dx$.
All demes are processed sequentially in a random order with the following scheme:
\begin{itemize}
	\item draw $B =  \text{Binomial}(N_1, p = D_1 \mathop{dt}/\mathop{dx}^2)$
	\item draw $B_\text{left} =  \text{Binomial}(B, p = 0.5)$
	\item update $N_1(x,t+\mathop{dt}) \rightarrow N_1(x,t) - B$
	\item update $N_1(x-dx,t+\mathop{dt}) \rightarrow N_1(x-dx,t) + B_\text{left}$
	\item update $N_1(x+dx,t+\mathop{dt}) \rightarrow N_1(x+dx,t) + (B-B_\text{left})$
\end{itemize}
Diffusion is performed in a similar way for species 2.

In practice, we set $\mathop{dt}=1$ and $\mathop{dx}=1$ without loss of generality. We also keep $r_1$, $r_2$, $D_1$, and $D_2$
small enough to ensure that the probabilities in binomial distributions remain smaller than one. 

\subsection*{Evolutionary simulations without trade-off}
At each time point, each division gives rise to a mutation, with a probability of 5\%. Each new phenotype are randomly drawn from a normal distribution centered on the ancestor phenotype, with a standard deviation of 0.1 times the phenotype in each direction. To speed up the simulations, subpopulations that did not reach a certain size after a certain time since they appeared are cleared up and their counts are randomly distributed over the remaining populations.

\subsection*{Evolutionary simulations with trade-off}
At each time point, each division gives rise to a mutation, with a probability of 1\%. Each new phenotype randomly falls on the trade-off line (uniform distribution), which is splitted into 500 bins within the first quadrant of the space ($r,D$).

\newpage{}
\renewcommand{\theequation}{D\arabic{equation}}
\renewcommand{\thetable}{D\arabic{table}}
\renewcommand{\thefigure}{D\arabic{figure}}
\setcounter{equation}{0}  
\setcounter{table}{0}
\setcounter{figure}{0}

\section*{Appendix D: Statistical analysis}
\subsection*{Figure 3C}
To study the shape of the four classes of outcomes in the density-distance space (Figure 3C), we performed a classification 
with a multinomial logistic regression. Each boundary between two adjacent classes is tested independantly.
In practice, we used the function \textit{mnrfit} of MATLAB to fit the data with a power law model. 
\begin{equation}
Class = a_0 + a_1 \log(Density) + a_2 \log(Distance)
\end{equation}
where \textit{Class} is 0 or 1 for each of the two tested classes. 

The boundary between the two classes is defined by $Class=0.5$:
\begin{equation}
Distance = \exp(\frac{0.5 - a_0}{a_2})Density^{-a_1/a_2}
\end{equation}
The exponent ($-a_1/a_2$) is lower than 1 (with statistical significance reported in the table below, the brackets represent 95\% confidence intervals which are
calculated from standard deviations calculated by the multinomial logistic regression and combined using Fieller's theorem) in the three cases indicating that boundaries between classes have a concave shape, in agreement with the 2-species FKPP model.

\begin{table}[h!]
	\centering
	\begin{tabular}{c c  }
		Boundary & exponent [95\% confidence interval] \\ \hline
		no trace - few traces &  0.24 [0.11, 0.38]\\
		few traces - partial sweep  & 0.43 [0.22, 0.63]\\
		partial sweep - full sweep & 0.57 [0.27, 0.88]
	\end{tabular}
	\label{table:stat}
\end{table}

\subsection*{Figure S2}
Figure S2 shows the results of the competition wildtype vs hyperswarmer on a plate when the spatial structure is suppressed. The wildtype outcompetes the hyperswarmer mutant. The significance of this result is estimated using a generalized linear model for binomial data (function \textit{fitglm.m} in MATLAB). We used the formula $c \sim 1 + f + (1 | R) + (f | S)$, where $c$ is the wildtype count, $f$ is the categorical variable 'before competition' or 'after competition', $R$ is the replicate index, and $S$ is the color index (we performed the two types of experiments: wildtype GFP vs. hyperswarmer DsRed, and wildtype DsRed vs. hyperswarmers GFP). The fit gave a $\text{p-value}=5.8 \times 10^{-5}$.







\newpage{}

\bibliographystyle{amnatnat}
\bibliography{AmNat.bib}

\newpage{}

\section*{Tables}
\renewcommand{\thetable}{\arabic{table}}
\setcounter{table}{0}

\begin{table}[h]
	\caption{Experimental swarming traits of \textit{P. aeruginosa} and its hyperswarmer mutant}
	\label{Table:traits}
	\centering
	\begin{tabular}{c c c c c}\hline
		Strain    & Growth rate $r$ & Expansion rate $v$  & Diffusion rate $D$  & Decay length $\lambda$  \\ 
		&  (measured in & (measured in & (calculated) & (calculated) \\
		& \citet{Deforet2015}) & \citet{Deforet2014})  & & \\
		\hline
		Wild-type  & $1.1  \pm 0.05 \text{h}^{-1}$  & $3.0 \pm 0.1 \text{mm}/\text{h}$ & $2.0 \pm 0.15 \text{mm}^2/\text{h}$ &  $1.4 \pm 0.07 \text{mm}$ \\
		Hyperswarmer   & $1.0 \pm 0.04 \text{h}^{-1}$ & $4.0 \pm 0.15 \text{mm}/\text{h}$  & $4.0 \pm 0.27 \text{mm}^2/\text{h}$ & $2.0 \pm 0.11 \text{mm}$ \\
	\end{tabular}
	\bigskip{}
	\\
	{\footnotesize Note: Errors are standard deviations. Errors on calculated parameters are inferred from the formula of propagation of errors.}
\end{table}

\newpage{}

\section*{Figure legends}
\renewcommand{\thefigure}{\arabic{figure}}
\setcounter{figure}{0}

\renewcommand{\thefigure}{A\arabic{figure}}
\setcounter{figure}{0}

\renewcommand{\thefigure}{B\arabic{figure}}
\setcounter{figure}{0}



\renewcommand{\figurename}{Video} 
\renewcommand{\thefigure}{\arabic{figure}}
\setcounter{figure}{0}

\begin{figure}[hp]
	\centering
	\begin{center}
		\includegraphics[width=6cm]{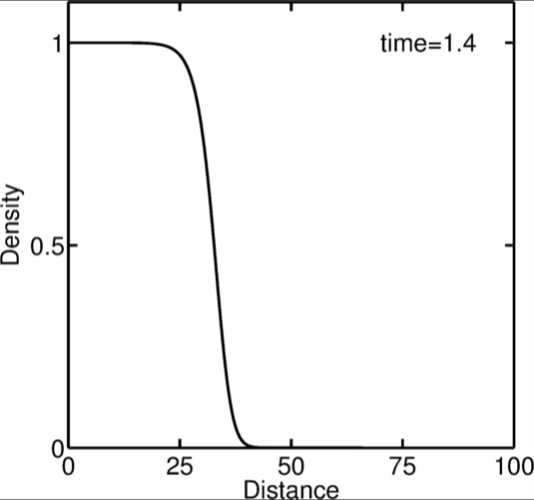}
	\end{center}
	\caption{Traveling wave solution of the F-KPP equation, advancing at the speed $v=2\sqrt(rD)$. }
\end{figure}

\begin{figure}[hp]
	\centering
	\begin{center}
		\includegraphics[width=10cm]{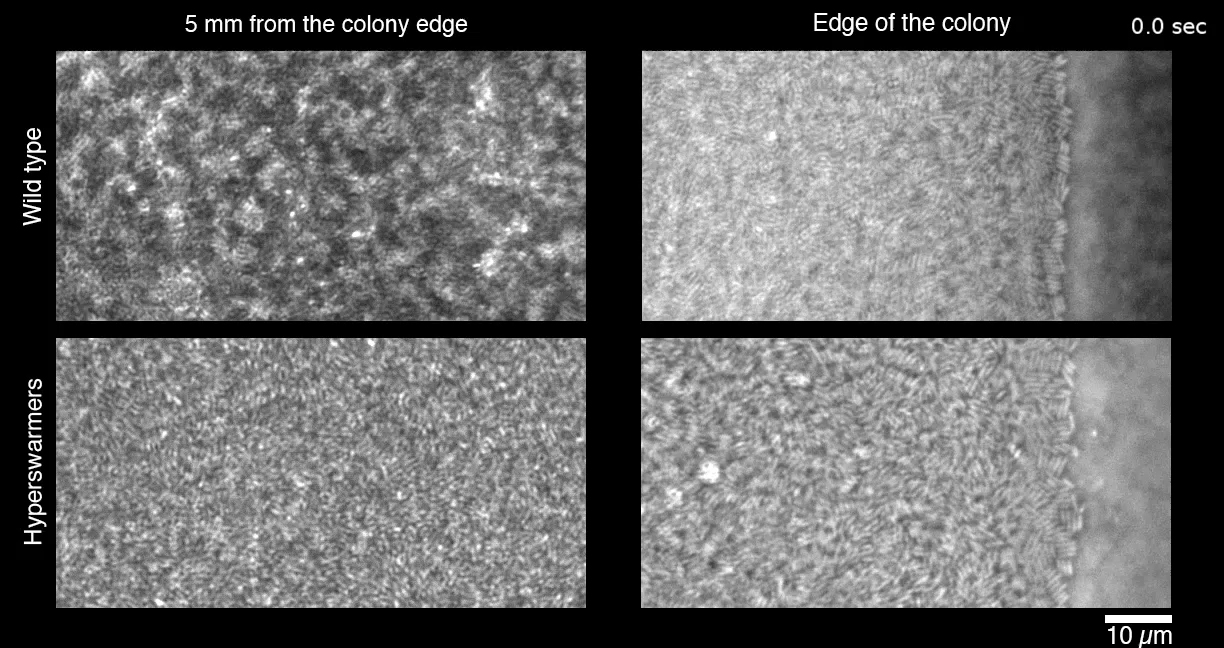}
	\end{center}
	\caption{Phase-contrast video-microscopy images of \textit{P. aeruginosa} swarming colonies. Top row: \textit{P. aeruginosa} wild-type colony. Bottom row: Hyperswarmer mutant colony. Left side: 5 mm from the edge. Right side: edge of the swarming colony. Note the active turbulence patterns displayed by the hyperswarmer colony, whereas wild-type cells seem more static. }
\end{figure}

\begin{figure}[hp]
	\centering
	\begin{center}
		\includegraphics[width=8cm]{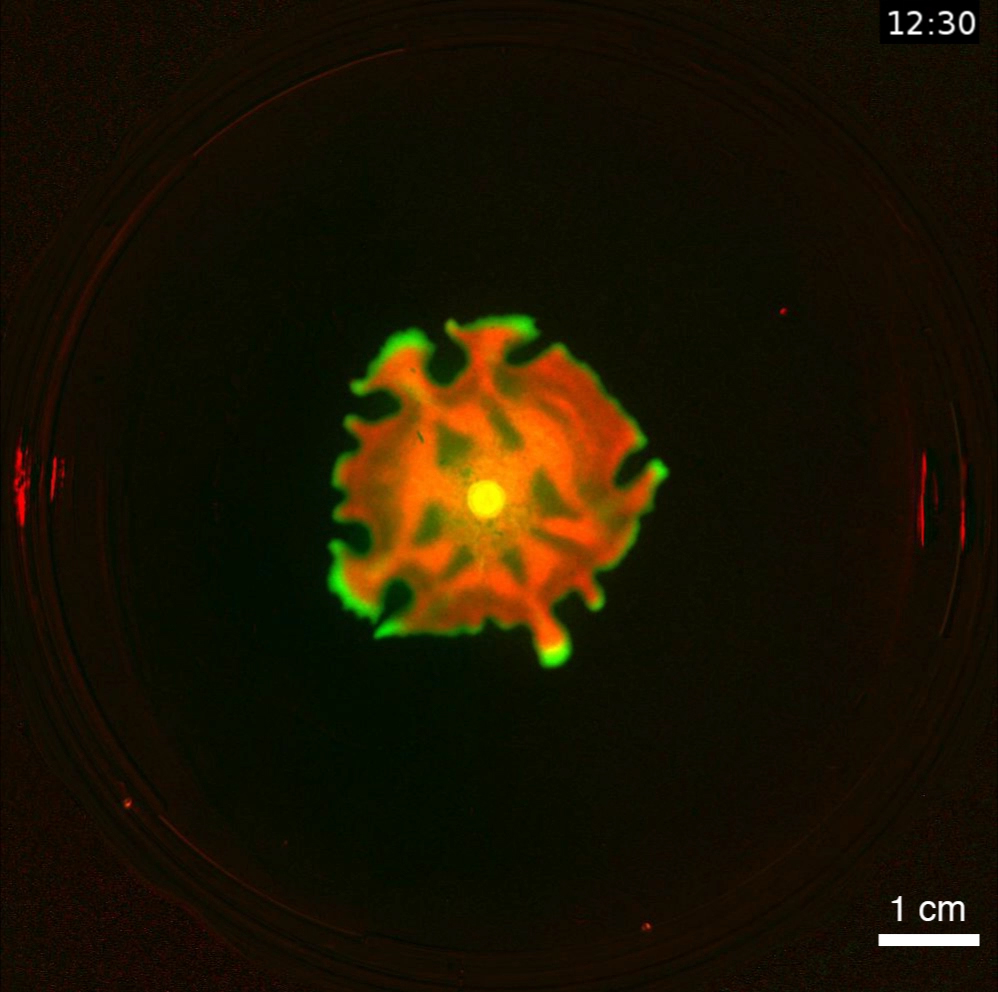}
	\end{center}
	\caption{ Fluorescence video of a swarming colony formed by a mixed population \textit{P. aeruginosa} wild type / Hyperswarmers (initial ratio 10:1).
		\textit{P. aeruginosa} wild-type constitutively expresses DsRed (red). Hyperswarmer mutant constitutively expresses GFP (green).}
\end{figure}

\begin{figure}[hp]
	\centering
	\begin{center}
		\includegraphics[width=12cm]{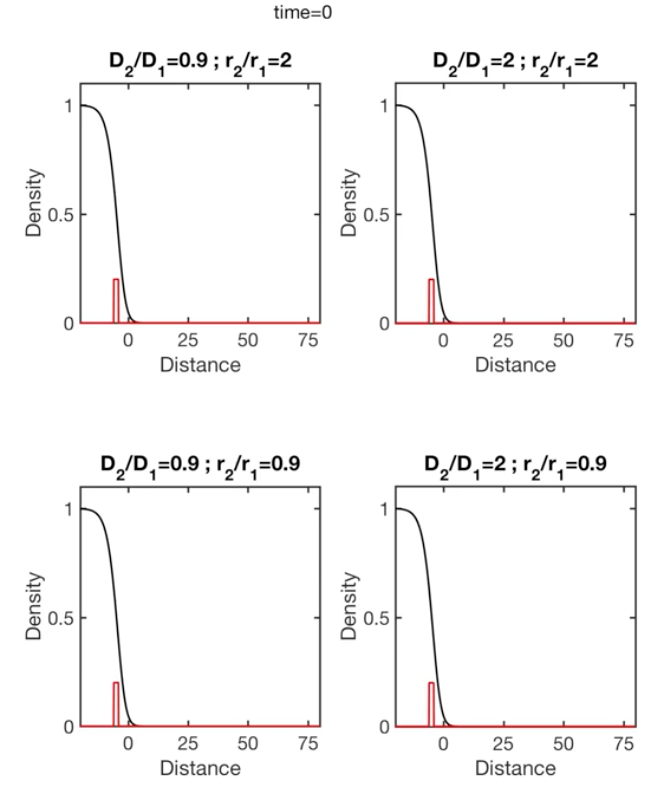}
	\end{center}
	\caption{Introduction experiments performed in numerical simulations. The black line represents the density of species 1. The red line represents species 2, implanted at t=0 at the distance L=5 from the edge.  }
\end{figure}

\begin{figure}[hp]
	\centering
	\begin{center}
		\includegraphics[width=12cm]{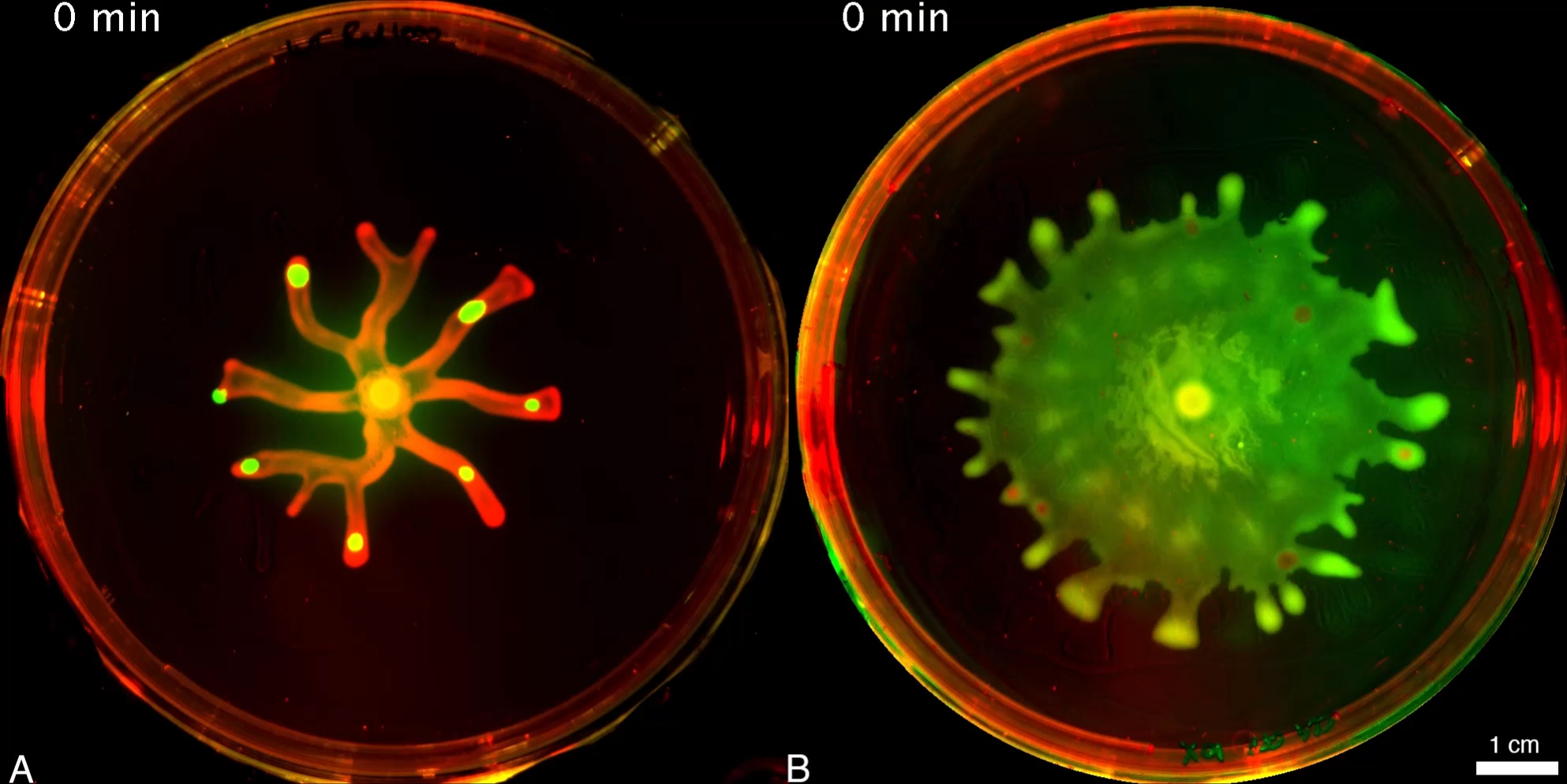}
	\end{center}
	\caption{Implantation experiments. A: Hyperswarmers (green) are implanted into wild type (red) swarming branches. B: Wild type (red) are implanted into a hyperswarmer colony.}
\end{figure}

\renewcommand{\figurename}{Figure}
\setcounter{figure}{1}


\subsection*{Online figure legends}
\renewcommand{\figurename}{Figure}
\renewcommand{\thefigure}{S\arabic{figure}}
\setcounter{figure}{0}

\begin{figure}[hp]
	\centering
	\begin{center}
		\includegraphics[width=13cm]{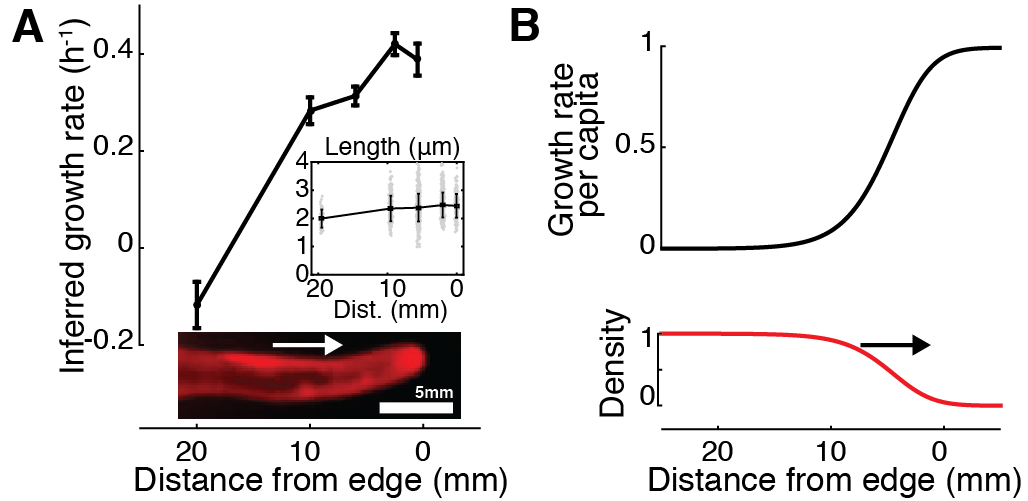}
		\end{center}
	\caption{Bacterial swarming can be modeled by FKPP equation where growth is greatest at the edge.
		A: The length of each bacterium is converted into a growth rate, using the results a previous study (\citealt{Deforet2015}) linking cell length to growth rate, where this fit is obtained: $L = 2.18 + 0.89\mu$ ($L$ is the length of a bacterium and $\mu$ is its growth rate in casa-aminoacids). Errors bars are standard error of the mean ($N \simeq 100$).
		Inset: Cells length measured by fluorescence microscopy and automated image analysis. Error bars are standard deviations. Data points are randomly distributed around each value of distance for better visualization. The picture represents a wildtype branch growing on agar gel. The white arrow depicts the direction of expansion.
		B: The growth rate per capita ($r(1-u)$, with $r=1$, black line), and the density ($u$) of the population simulated by the FKPP equation (black line). The black arrow depicts the direction of expansion.}
	    \label{fig:figS1}
\end{figure}

\begin{figure}[hp]
	\centering
	\begin{center}
		\includegraphics[width=16.5cm]{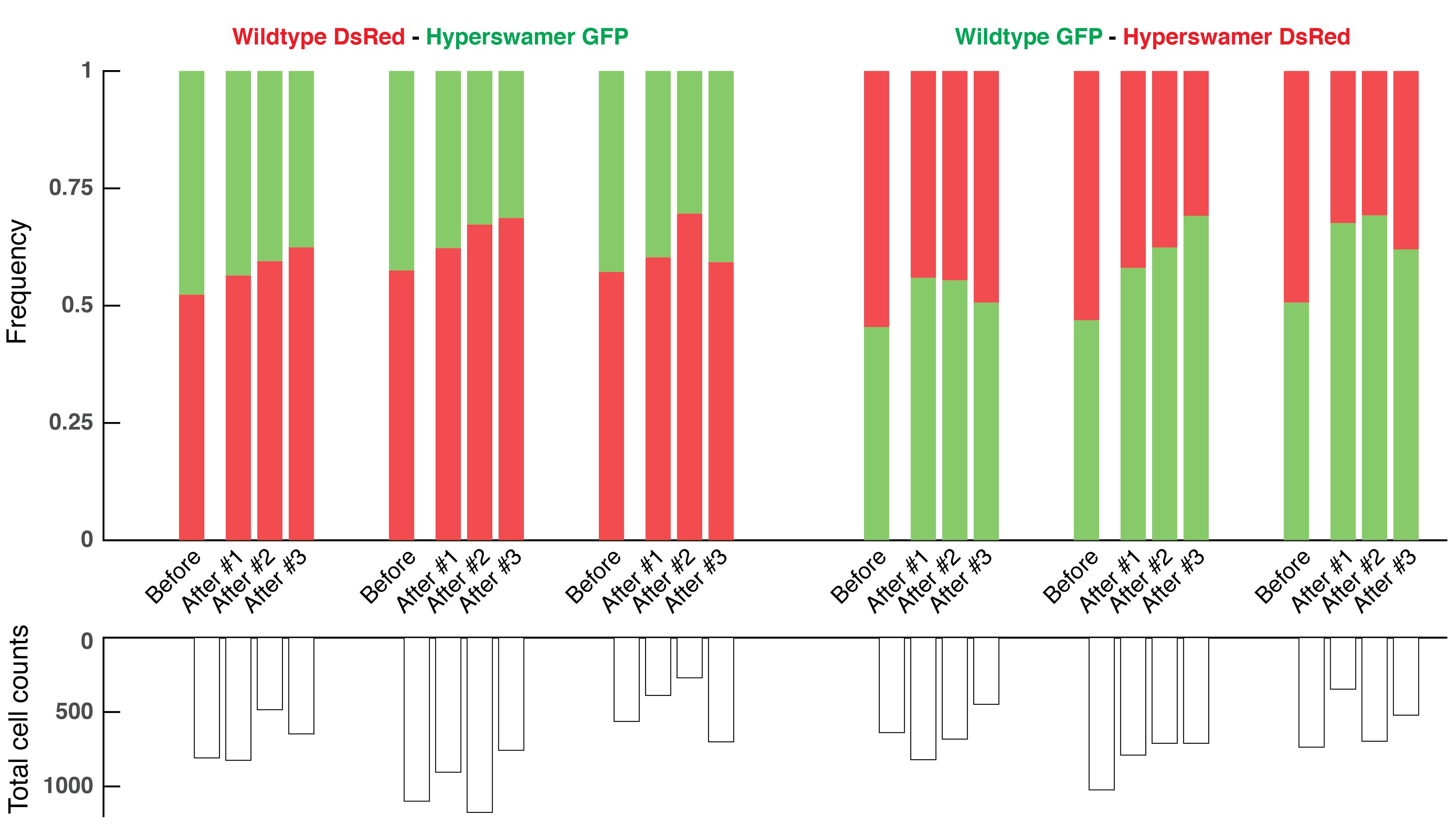}
	\end{center}
	\caption{Wild type \textit{P. aeruginosa} outcompetes its hyperswarmer mutant on a plate when the spatial structure is suppressed.
		Swarming plates (0.5\% agar) are inoculated with a loan of a mixed population of wildtype and hyperswarmers (approximate initial ratio 1:1) and incubated for 4 hours of incubation at 37\textdegree C. The ratios of wild type before and after the competition are estimated using CFUs counting. GFP and DsRed fluorescent proteins (constitutively expressed) are used to take apart the two bacterial strains.  Swapped experiments confirm that despite toxicity of DsRed proteins, wild type cells outcompete hyperswarmers ($p=5.8\times 10^{-5}$, generalized linear model for binomial data). N=3 biological replicates x 3 technical replicates per color combination.}
	\label{fig:figS2}
\end{figure}

\begin{figure}[hp]
	\centering
	\begin{center}
		\includegraphics[width=10cm]{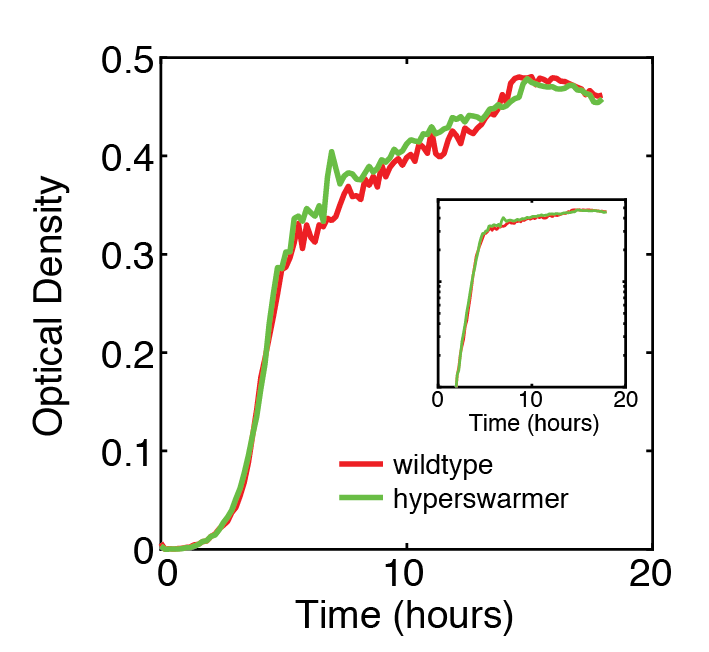}
	\end{center}
	\caption{Wild type \textit{P. aeruginosa} and its hyperswarmer mutant have comparable carrying capacities.
		The bacterial density is measured in a plate scanner as optical density. Inset: Growth curves in logarithmic scale.}
	\label{fig:figS3}
\end{figure}

\begin{figure}[hp]
	\centering
	\begin{center}
		\includegraphics[width=17cm]{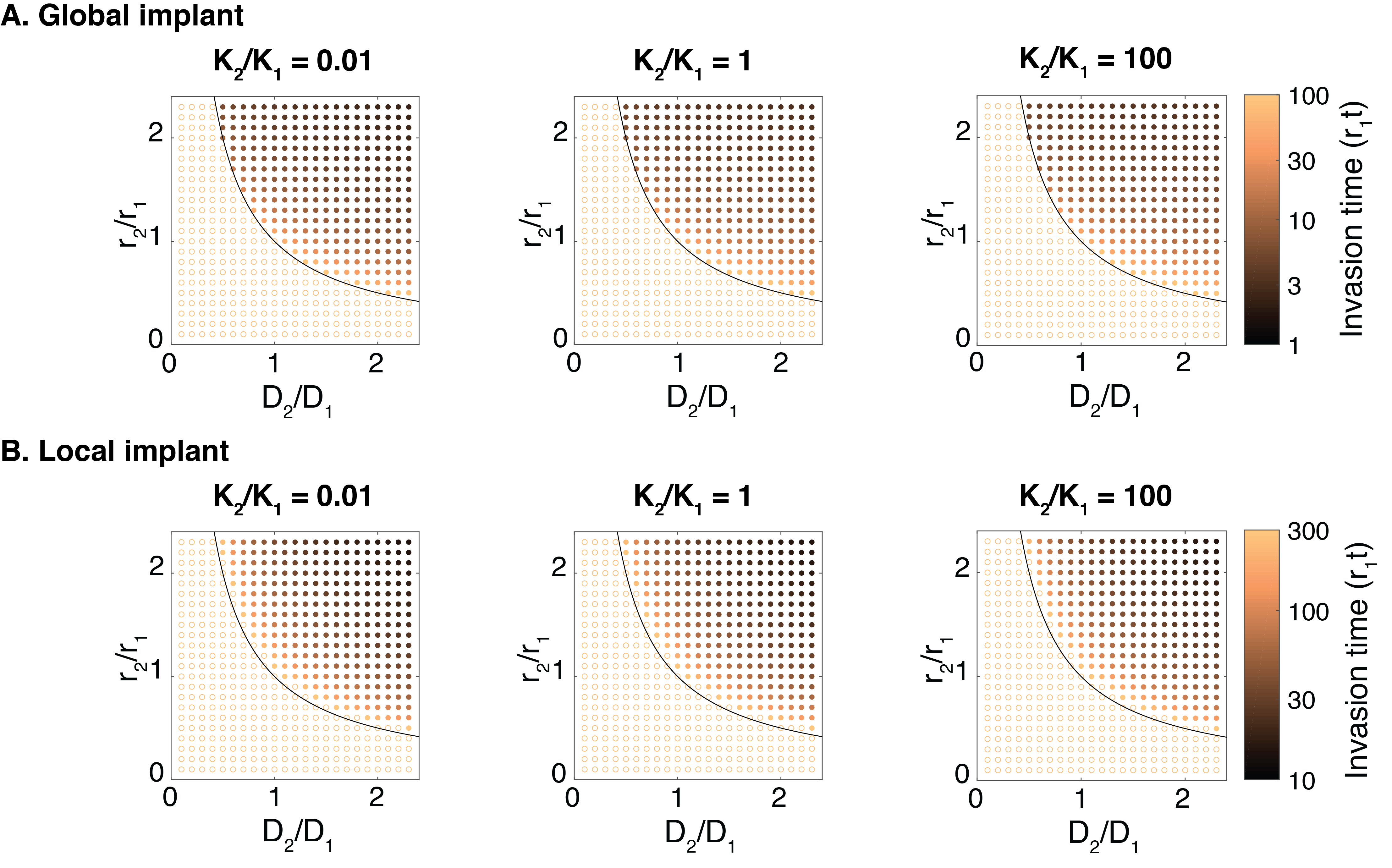}
	\end{center}
	\caption{Different carrying capacities do not affect the evolutionary processes.
		A: Species 2 is introduced globally with $u_2=u_1/100$. B: Species 2 is introduced locally with $u_2(x)=0.2$ where $-16<x<-14$ (with the edge of the population being at $x=0$).}
	\label{fig:figS4}
\end{figure}

\begin{figure}[hp]
	\centering
	\begin{center}
		\includegraphics[width=17cm]{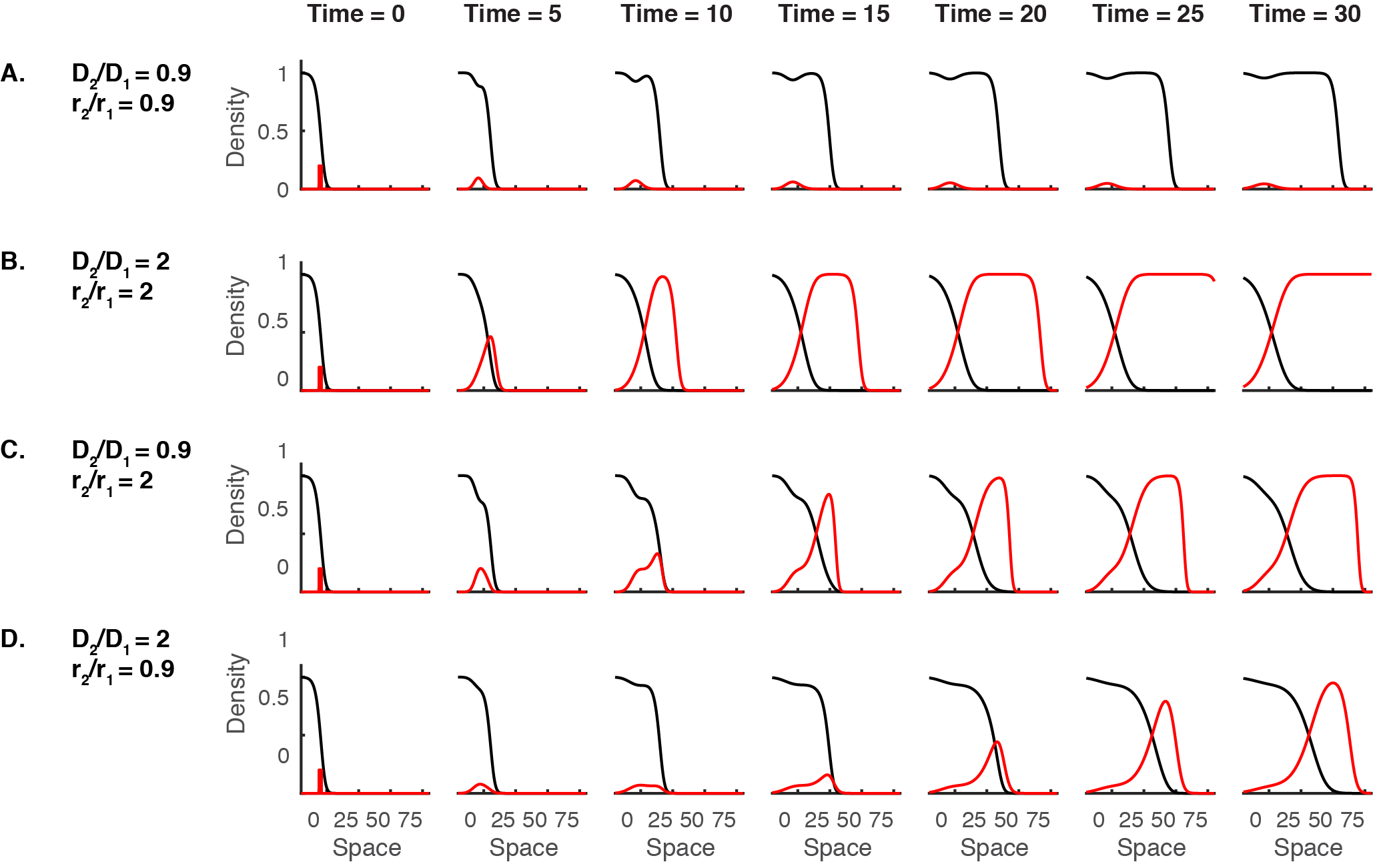}
	\end{center}
	\caption{Snapshots of takeover (or no takeover) after introduction of species 2.
		A: No takeover case with lower growth and lower dispersal.
		B: Takeover case with greater growth and greater dispersal.
		C: Takeover case with greater growth but lower dispersal.
		D: Takeover case with lower growth but greater dispersal. See also Video 4.}
	\label{fig:figS5}
\end{figure}

\begin{figure}[hp]
	\centering
	\begin{center}
		\includegraphics[width=17cm]{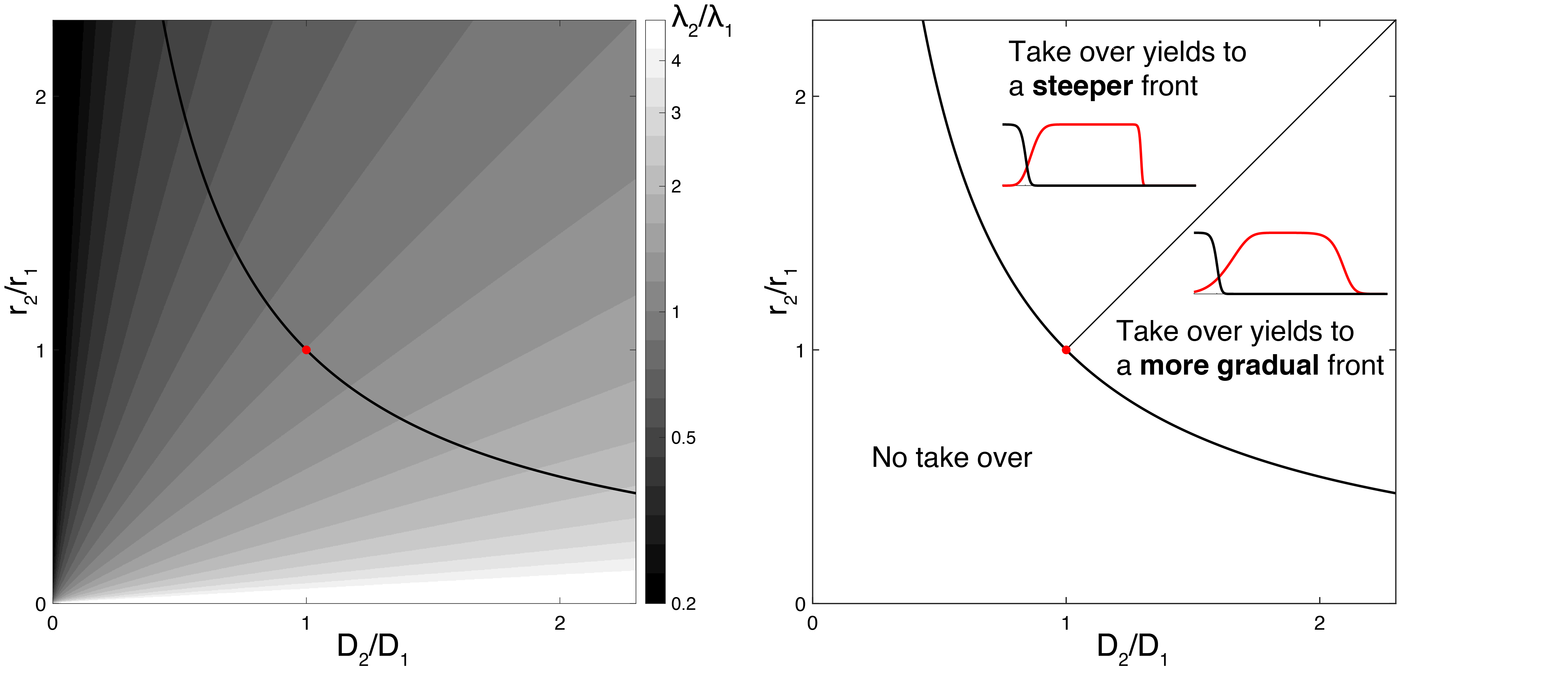}	
	\end{center}
	\caption{The lengthscale of the front is modified when another phenotype takes over.
		The lengthscale of the front ($\lambda=\sqrt{D/r}$) depends on the phenotype of the population sitting at the front. Here is a map of the change of lengthscale ($\lambda_2/\lambda_1$) in the space ($r$,$D$). The red dot depicts the phenotype of the ancestral (reference) species. The black line represents the success rule. Note the chosen colormap is not smooth to highlight the shape of the map. Right panel: a chart representing domains of takeover with a steeper front and with a more gradual front.}
	\label{fig:figS6}
\end{figure}

\begin{figure}[p]
	\centering
	\begin{center}
		\includegraphics[width=10cm]{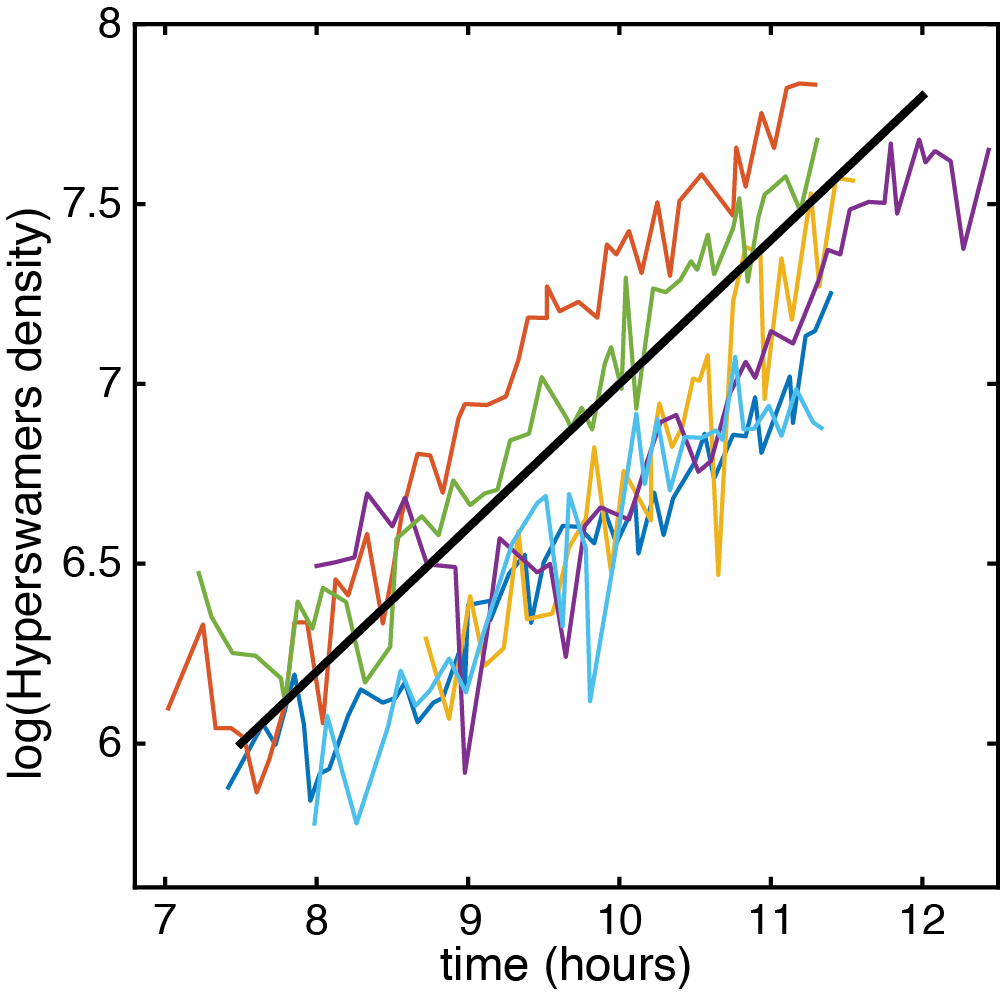}
	\end{center}
	\caption{Hyperswarmers density dynamics at the edge of the colony, from Video 1. The dynamics of the density of hyperswarmers is governed by the eigenvalue $\tilde r _{\text{max}} = r_2 - \frac{v_1^2}{4D_2}$, in the limit of low density.
		Experimental parameters (Table \ref{Table:traits}) yields to $\tilde r _{\text{max}} = 0.4 \text{ h}^{-1}$. The black line is a visual guide with a rate of $0.4 \text{ h}^{-1}$. The average slope of the six curves is $0.39$ \textpm $0.08 \text{h}^{-1}$ (Standard deviation). }
	\label{fig:figS7}
\end{figure}

\begin{figure}[p]
	\centering
	\begin{center}
		\includegraphics[width=12cm]{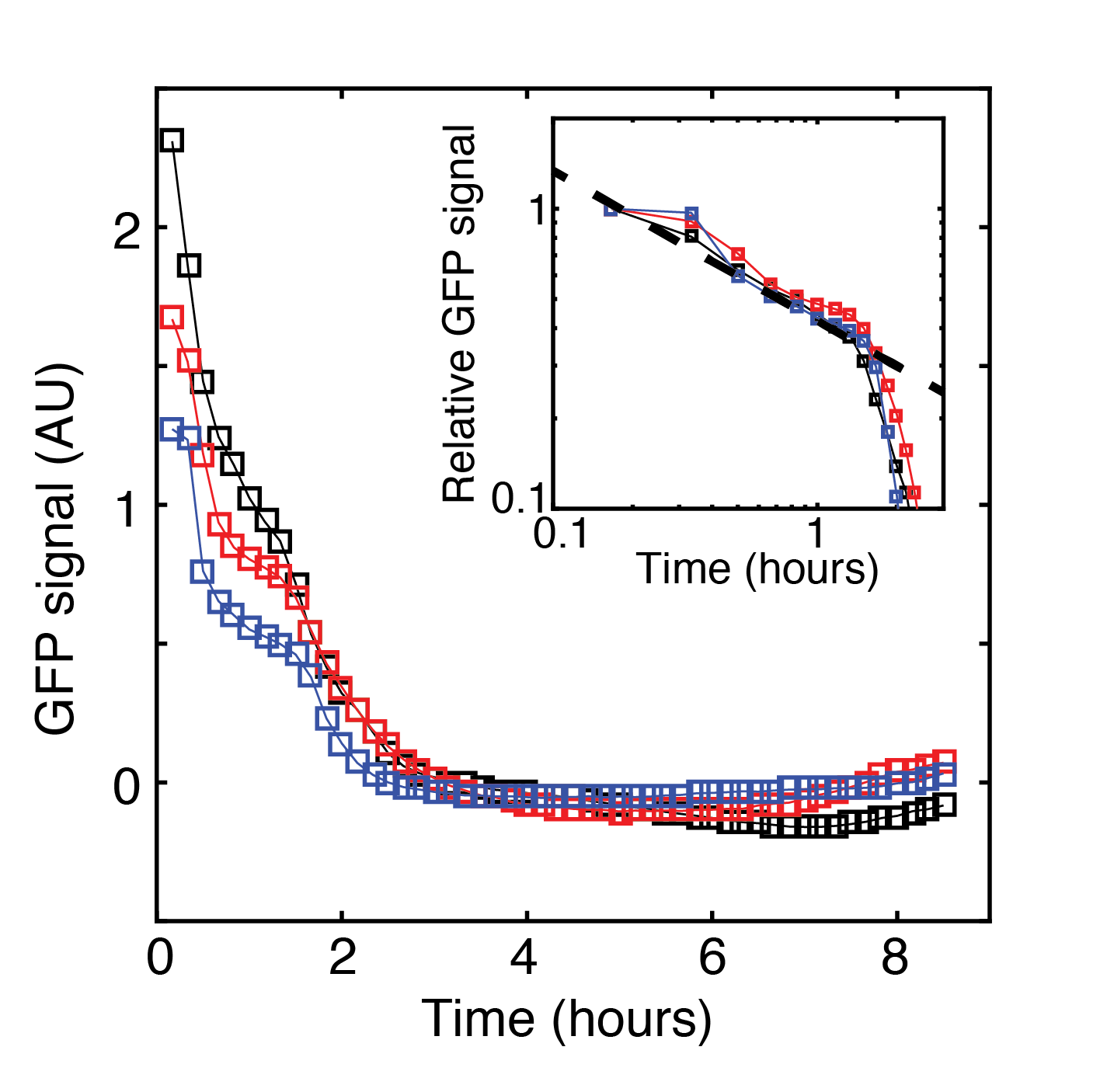}
	\end{center}
	\caption{Hyperswarmers introduced into a wildtype swarming colony spread by diffusion. Three representative implant sites are analyzed. The data represents the average GFP signal at the site of introduction. GFP signal 1 cm closer to the center of the colony is used for background correction. Data extracted from video 5A. Inset: the first 1.5 hours can be modeled by a diffusive decay. The dashed line is the prediction from 1D diffusion decay, using $D=4 \text{ mm}^2/\text{h}$ and a inoculum size of 3 mm ($C\sim \frac{C_\text{max}L}{\sqrt{4\pi D t}}$).}
	\label{fig:figS8}
\end{figure}

\begin{figure}[p]
	\centering
	\begin{center}
		\includegraphics[width=17cm]{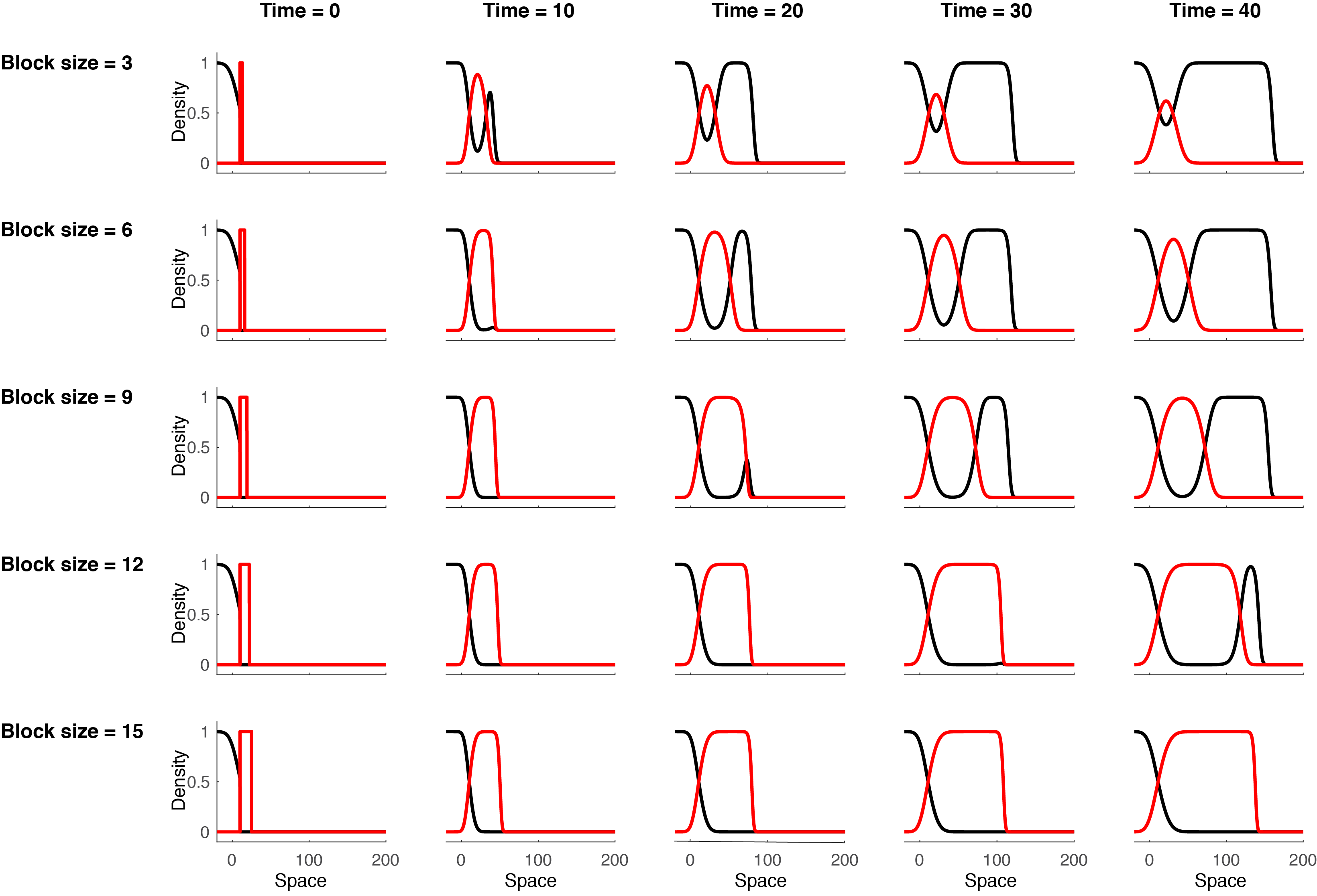}
		\end{center}
	\caption{Wild-type cannot prevent hyperswarmers from taking over. A $t=0$, a block of species 2 (red curve, $r_2=1.1$ and $D_2=0.5$) is implanted at the edge of the population of species 1 (black curve, $r_1=1$ and $D_1=1$). Formally, the simulation starts with the steady-state profile of species 1, where individuals are entirely removed at the edge ($u_1$ is unchanged for $x<10$, $u_1=0$ for $x>10$). For species 2, $u_2=1$ for $10<x<10+\textit{Block size}$ (with $\textit{Block size}$ from 3 to 15), and $u_2=0$ everywhere else.  A larger block size delays the takeover, but does not prevent it. Take over is nearly immediate for block size lower than 3. For larger blocks, species 2 forms a traveling wave for a short time before being taken over. For block size greater that 15, take over is not observed, even at longer simulation time, possibly because of numerical precision issues at very low population densities. Similar behavior is observed for all values of $r_2$ and $D_2$, as long as $v_2<v_1$.}
	\label{fig:figS9}
\end{figure}

\begin{figure}[p]
	\centering
	\begin{center}
		\includegraphics[width=12cm]{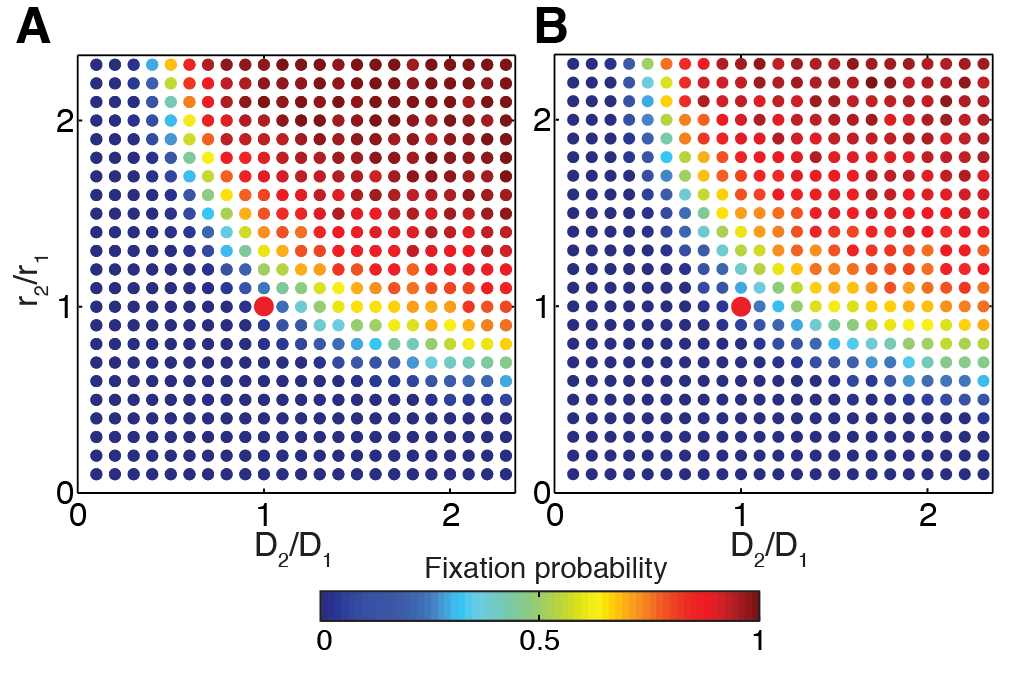}
	\end{center}
	\caption{Evolutionary diagram in stochastic models. Fixation probability obtained from the stochastic model without death rate (A), and with death rate (B). For the stochastic simulations, $S=1$, $K=100$, $L=2\lambda_1$ . The red dot depicts the reference point ($D_2=D_1$ and $r_2=r_1$).}
	\label{fig:figS10}
\end{figure}

	\begin{figure}[h!tp]
		\centering
		\begin{center}
			\includegraphics[width=14cm]{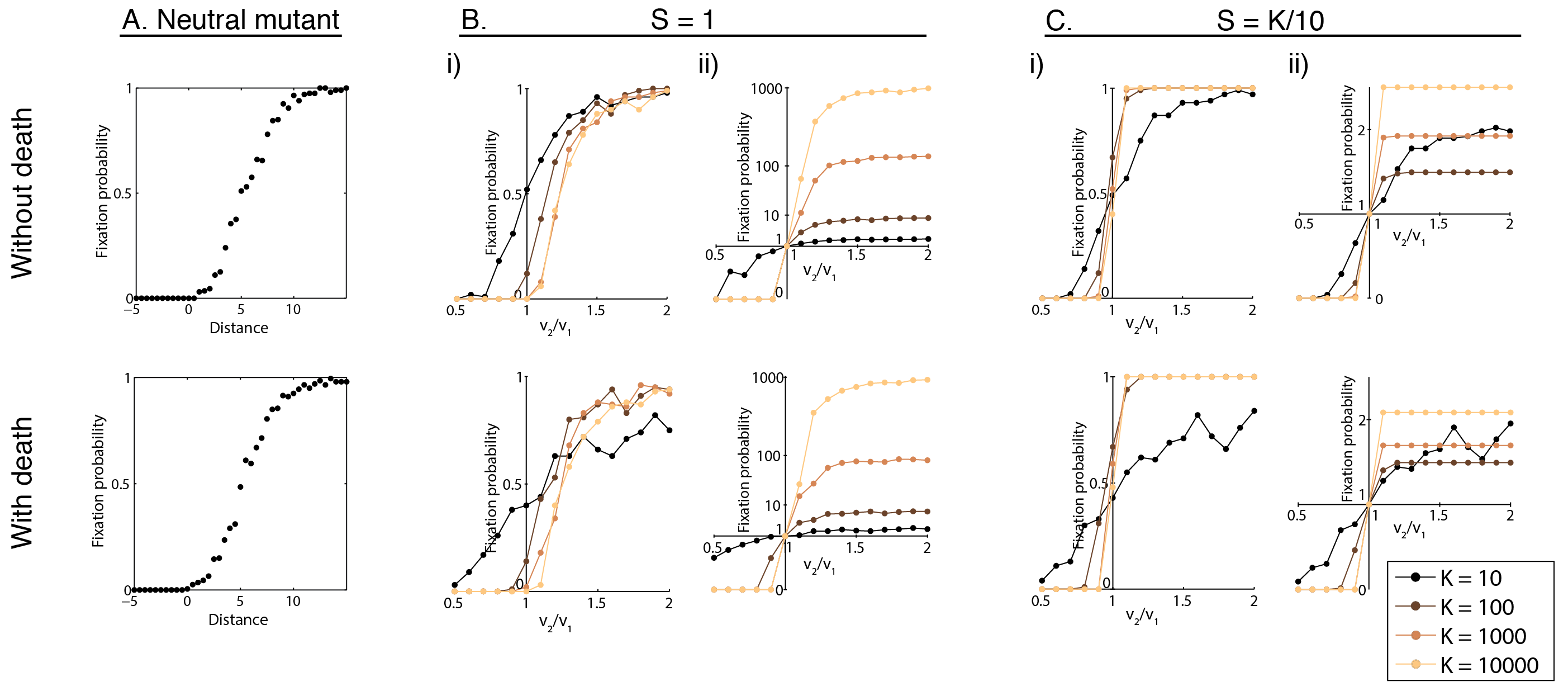}
		\end{center}
		\caption{Fixations probability in stochastic models.
			The first row shows results of the stochastic model without death. The second row shows results of the stochastic model with death.
			A. Fixation probability of a neutral mutant with respect to the distance to the front ($S=1$).
			B. Fixation probability of a mutant with respect to $v_2/v_1$ (with $r_2/r_1 = D_2/D_1$), for various carrying capacities. 
			The introduction occurs in fixed size implant ($S=1$ individual) at $L=3$.
			i) represents the raw data in linear scale. ii) represents the data normalized by the fixation probability of a neutral mutant (note power-law scale on the y-axis, exponent=0.2).
			C. Fixation probability of a mutant with respect to $v_2/v_1$ (with $r_2/r_1 = D_2/D_1$), for various carrying capacities. 
			Unlike in B, the number of introduced individuals scales with carrying capacity ($S=K/10$).
			i) represents the raw data in linear scale. ii) represents the data normalized by the fixation probability of a neutral mutant, in linear scale.
			Note that in B and C, for the model with and without death, the fixation probability curves get steeper as the carrying capacity gets larger.
		}
		
		\label{fig:stochastic}
	\end{figure}

\begin{figure}[h!tp]
	\centering
	\begin{center}
		\includegraphics[width=17cm]{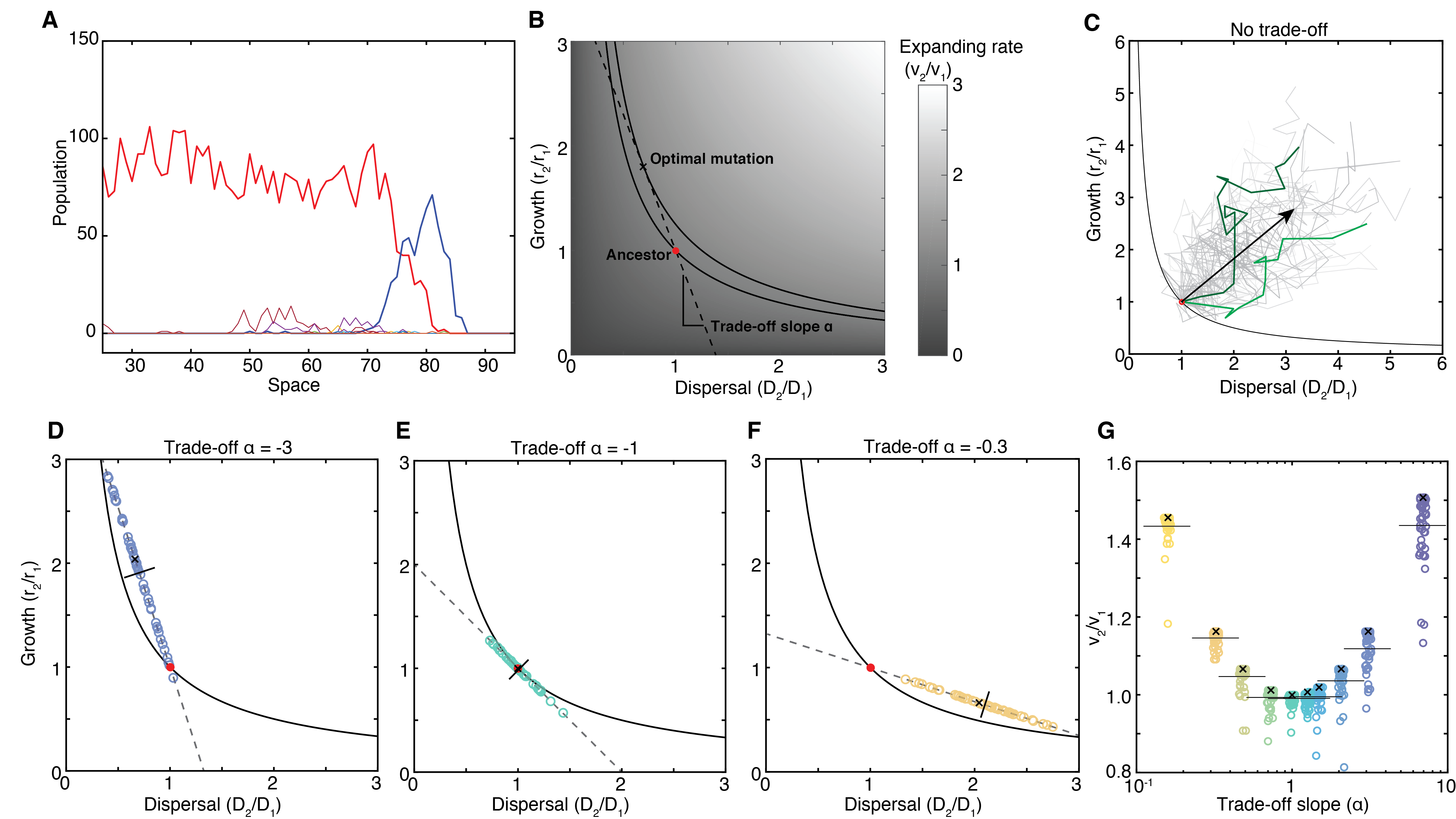}
	\end{center}
	\caption{Stochastic simulations of population expansion with mutations reveal that evolution tends to maximize the expansion rate $v=2\sqrt{rD}$. 
		A: Representative snapshot of the spatial distributions of a population (red) and mutant populations it has generated.
		B: Diagram representing the success rule (black line), the ancestral phenotype (red dot), a trade-off line (dashed line), the expansion rate (grey levels), and the phenotype that maximizes the expansion rate along the trade-off line (cross symbol).
		C: Without trade-off between growth and dispersal, evolution tends to improve both traits simultaneously. 50 evolutionary trajectories of the majority phenotype sitting at the edge are represented.  Two trajectories are highlighted in green. The black arrow depicts the average trajectory (average from 50 simulations).
		D-E-F: Phenotypes obtained after $r_{1}t=1000$ with 3 trade-off slopes: -0.5, -1, -2. The black line is the average phenotype from 50 simulations, the cross is the theoretical optimal phenotype. 
		G: Summary of the results for 10 trade-off slopes.
	}
	\label{fig:simuevo}
\end{figure}

\begin{figure}[h!tp]
	\centering
	\begin{center}
		\includegraphics[width=10.8cm]{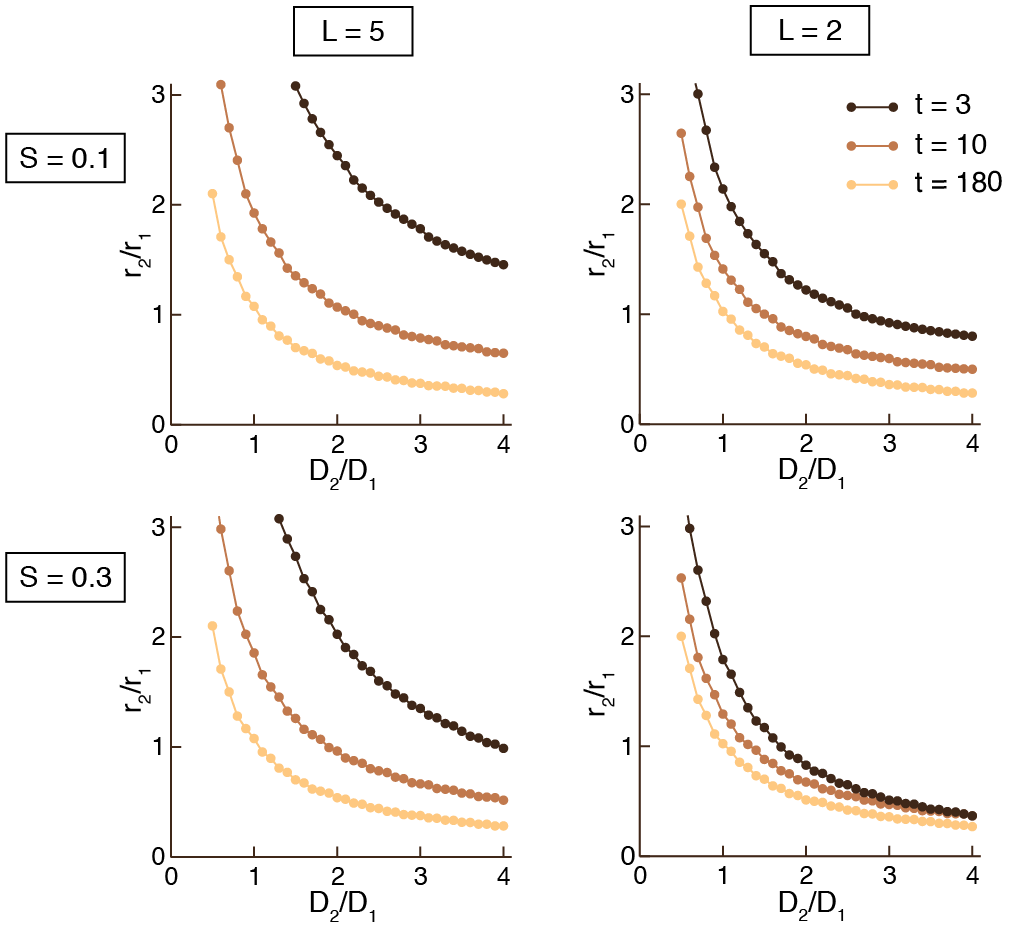}
	\end{center}
	\caption{Evolutionary domain boundaries for various simulation times, sizes of introduction $S$, and distances of introduction $L$.}
	\label{fig:phase}
\end{figure}

\begin{figure}[h!tp]
	\centering
	\begin{center}
		\includegraphics[width=10.8cm]{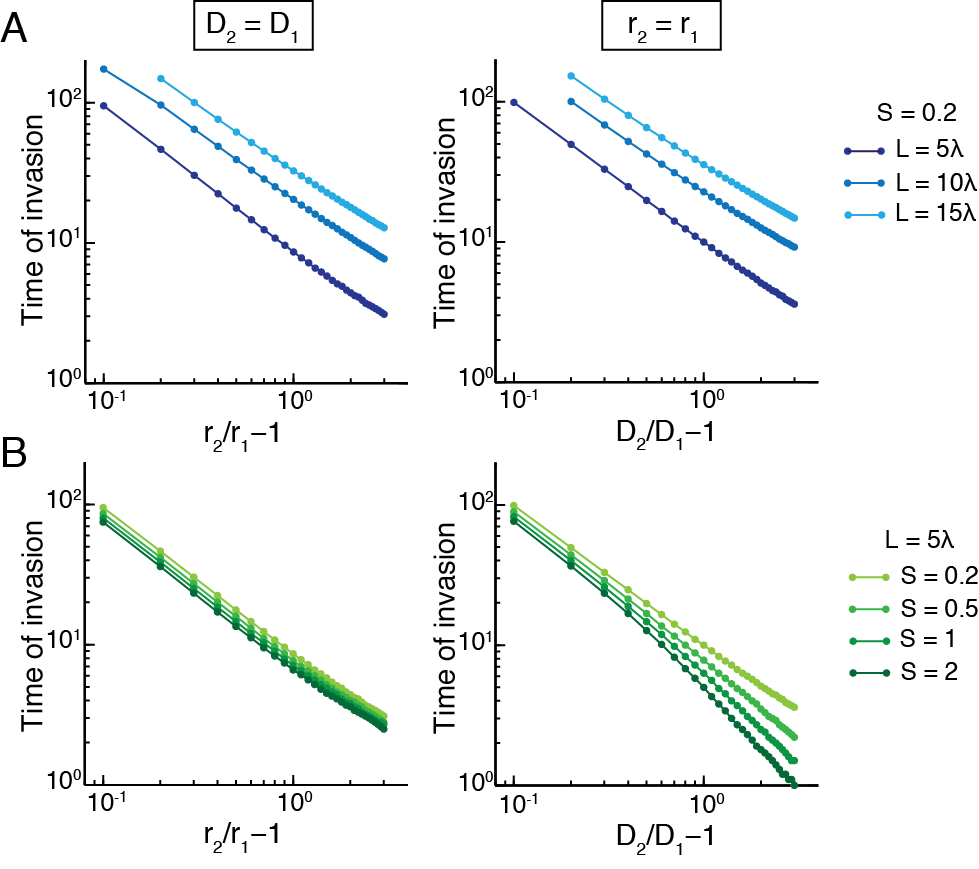}
	\end{center}
	\caption{Implanting further away or lower amounts delays takeover. A: Time of 
		takeover for different implantation sites. B: Time of takeover for different 
		implant sizes.}
	\label{fig:delay}
\end{figure}

\end{document}